\pdfoutput=1
\newcommand*{\ATLASLATEXPATH}{}
\documentclass[cernpreprint,texlive=2011,txfonts,UKenglish]{\ATLASLATEXPATH atlasdoc}

\usepackage{bm}
\usepackage{lineno}
\usepackage{units}
\usepackage{placeins}
\usepackage{multirow}

\def\Hp {\ensuremath{H^+}}
\def\Hm {\ensuremath{H^-}}
\def\Hplus {\ensuremath{\Hp}}

\def\Hpm {\ensuremath{H^{\pm}}}

\def\tanbeta {\ensuremath{\tan\beta}}
\def\tanb {\ensuremath{\tanbeta}}

\def\kt{\ensuremath{k_{t}}} 

\def\invfb {\ensuremath{{\mathrm{fb}^{-1}}}}

\def\fb {\ensuremath{{\mathrm{fb}}}}
\def\pb {\ensuremath{{\mathrm{pb}}}}
\def\nb {\ensuremath{{\mathrm{nb}}}}

\def\Wprime {\ensuremath{\Wboson^{\prime}}}

\def\qprime {\ensuremath{q^{\prime}}}

\usepackage{\ATLASLATEXPATH atlaspackage}
\usepackage{\ATLASLATEXPATH atlasbiblatex}

\usepackage{\ATLASLATEXPATH atlascontribute}

\usepackage{\ATLASLATEXPATH atlasphysics}

\graphicspath{{logos/}{}}

\AtlasTitle{Search for charged Higgs bosons in the $\Hpm \to tb$ decay channel
in $pp$ collisions at $\sqrt{s} = \unit[8]{\TeV}$ using the ATLAS detector}
\author{The ATLAS Collaboration}
\nolinenumbers
\AtlasRefCode{HIGG-2013-28}
\PreprintIdNumber{CERN-PH-EP-2015-290}

\AtlasAbstract{%
Charged Higgs bosons heavier than the top quark and decaying via $\Hpm \to tb$
are searched for in proton--proton collisions measured with the ATLAS experiment at
$\rts = \unit[8]{\TeV}$ corresponding to an integrated luminosity of
$\unit[20.3]{\ifb}$. The production of a charged Higgs boson in association
with a top quark, $gb \to t\Hpm$, is explored in the mass range 200 to 600~\GeV\
using multi-jet final states with one electron or muon. In order to separate
the signal from the Standard Model background, analysis techniques combining
several kinematic variables are employed. %
An excess of events above the background-only hypothesis is observed across a wide mass range, amounting to up to 2.4 standard deviations.
Upper limits are set on the $gb\rightarrow tH^{\pm}$ production cross section times the branching
fraction $\mbox{BR}(\Hpm \to tb)$.
Additionally, the complementary $s$-channel production, $q\qprime \to \Hpm$, is
investigated through a reinterpretation of $\Wprime \to tb$ searches in ATLAS.
Final states with one electron or muon are relevant for $\Hpm$ masses from 0.4
to 2.0~\TeV, whereas the all-hadronic final state covers the range 1.5 to 3.0~\TeV.
In these search channels, no significant excesses from the predictions of the Standard Model are observed,
and upper limits are placed on the $q\qprime \to \Hpm$ production cross section times the branching
fraction $\mbox{BR}(\Hpm \to tb)$.}

\AtlasJournal{JHEP}

\AtlasCoverSupportingNote{Top-quark associated production}{https://cds.cern.ch/record/1637456}
\AtlasCoverSupportingNote{$s$-channel analysis (all hadronic)}{https://cds.cern.ch/record/1968160}
\AtlasCoverSupportingNote{$s$-channel analysis (lepton+jets)}{https://cds.cern.ch/record/1971086}

\AtlasCoverCommentsDeadline{}

\AtlasCoverAnalysisTeam{Liron Barak, Elin Bergeaas Kuutmann, Hadar Cohen, Julien Donini,
Johannes Erdmann, Erez Etzion, Arnaud Ferrari, Joakim Gradin, Geoffrey Gilles, Eilam Gross, Geoffrey Herbert,
Alexander Khanov, Andrea Knue, Ho Ling Li, Jeremy Love, Alexander Madsen, Kazuki Nagata, Gabriela Navarro,
Andrei Nikiforov, Daniel Pelikan, Michael Pitt, Silje Raddum, Flera Rizatdinova, Carlos Sandoval, Koji Sato,
Jana Schaarschmidt, Dimitri Sidorov, Timoth\'ee Th\'eveneaux-Pelzer, Jie Yu, Martin zur Nedden}

\AtlasCoverEdBoardMember{James Ferrando (chair), Fabrizio Parodi, Pippa Wells, Norbert Wermes}

\AtlasCoverEgroupEditors{atlas-HIGG-2013-28-editors@cern.ch}

\AtlasCoverEgroupEdBoard{atlas-HIGG-2013-28-editorial-board@cern.ch}

\hypersetup{pdftitle={ATLAS draft},pdfauthor={The ATLAS Collaboration}}

\begin{document}

\maketitle

\tableofcontents

\section{Introduction}
\label{sec:Introduction}

The discovery of a neutral scalar particle $H$ at the Large Hadron Collider (LHC)
in 2012~\cite{Aad:2012tfa,Chatrchyan:2012ufa}, with a measured mass of
$125.09 \pm 0.21 (\mbox{stat.}) \pm 0.11 (\mbox{syst.})$~\GeV~\cite{combinedHmass},
raises the question of whether this new particle is the Higgs boson of the Standard
Model (SM) or one physical state of an extended Higgs sector. The observation of
a heavy charged scalar particle would clearly indicate physics beyond the SM.
Charged Higgs bosons\footnote{In the following, charged Higgs bosons are denoted $\Hp$, with the charge-conjugate $\Hm$ always implied. Similarly, generic quark symbols are used for $q$ and $\bar{q}$.}
are predicted by several non-minimal Higgs scenarios, such as two-Higgs-doublet Models
(2HDM)~\cite{2hdm} and models containing Higgs triplets~\cite{trip1,trip2,trip3,trip4,trip5}.

\begin{figure}
\begin{center}
\includegraphics[height=3.6cm]{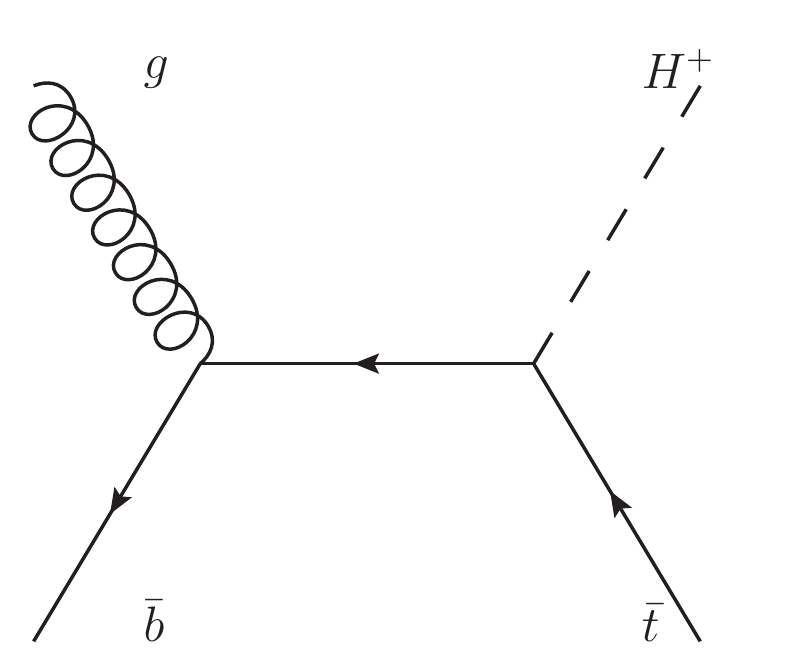}
\includegraphics[height=3.6cm]{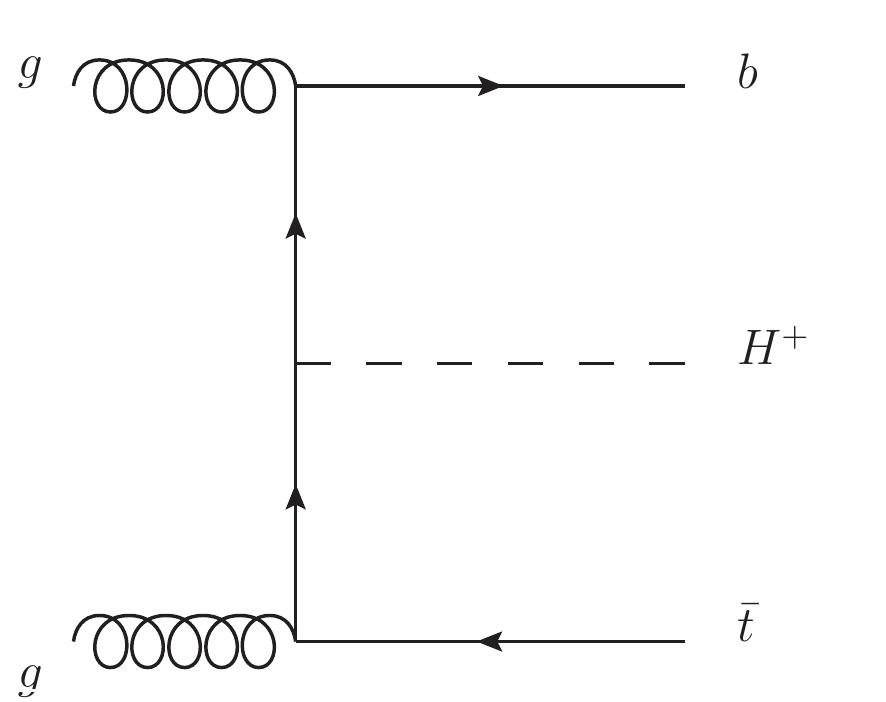}
\includegraphics[height=3.6cm]{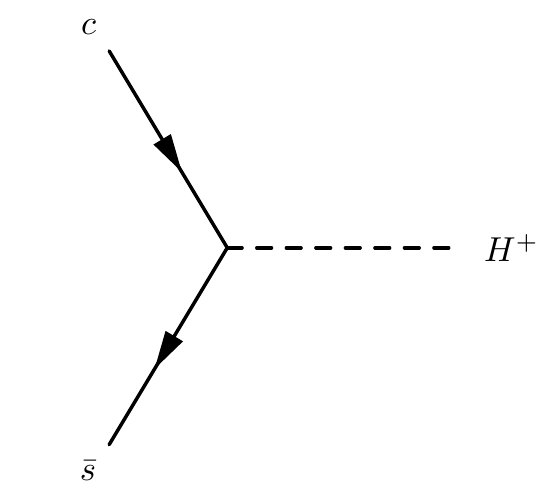}
\end{center}
\caption{Leading-order Feynman diagrams for the production of a charged Higgs boson with
a mass $m_{\Hp}>m_{\text{top}}$, in association with a top quark (left in the 5FS, and
centre in the 4FS) and in the $s$-channel (right). \label{fig:hplus_feynman_production}}
\end{figure}

The production mechanisms and decay modes of a charged Higgs boson depend on its
mass, $m_{\Hp}$. For light charged Higgs bosons ($m_{\Hp} \lesssim m_{\text{top}}$, where
$m_{\text{top}}$ is the top-quark mass), the primary production mechanism is
through the decay of a top quark, $t \to b\Hp$. For $m_{\Hp} > m_{\text{top}}$,
the dominant $\Hp$ production mode at the LHC is expected to be in association with a top quark,
as illustrated by the left-hand and central plots of figure~\ref{fig:hplus_feynman_production}.
When calculating the corresponding cross section in a four-flavour scheme (4FS),
$b$-quarks are dynamically produced, whereas in a five-flavour scheme (5FS), the
$b$-quark is also considered as an active flavour inside the proton. The 4FS and 5FS
cross sections are averaged according to ref.~\cite{Harlander:2011aa}.
In the 2HDM, the production and decay of the charged Higgs boson also depend on
the parameter $\tanb$, defined as the ratio of the vacuum expectation values of
the two Higgs doublets, and the mixing angle $\alpha$ between the CP-even Higgs
bosons. For $m_{\Hp} > m_{\text{top}}$ and in the case of
$\cos(\beta - \alpha)\approx 0$, the dominant decay is $\Hp \to tb$,
with a substantial contribution from $\Hp \to \tau\nu$ for large
values of $\tanb$~\cite{xsec2}.
A complementary \Hp\ production mode, shown in the right-hand plot of
figure~\ref{fig:hplus_feynman_production}, is the $s$-channel process,
$q\qprime \to \Hplus$.

The LEP experiments placed upper limits on the production of $H^{+}$ in the mass range of
40--100~\GeV~\cite{Abbiendi:2013hk}, and the Tevatron experiments set upper limits on
$\mbox{BR}(t \rightarrow bH^{+})$ for $m_{\Hp}$ in the range 80--150~\GeV~\cite{cdfRef,dzeroRef}.
The D0 experiment also searched for a charged Higgs boson with a mass in the range 180--300~\GeV\ using the $H^+ \rightarrow tb$ decay channel~\cite{Abazov:2008rn}.
Light charged Higgs bosons have been searched for in the $\tau\nu$ decay mode at the LHC by CMS (2~\invfb, $\sqrt{s} = 7$~\TeV~\cite{:2012cwb}) and ATLAS (4.7~\invfb, $\sqrt{s} = 7$~\TeV~\cite{Aad:2012tj,Aad:2012lea}). Searches for charged Higgs bosons were also performed in proton--proton ($pp$) collisions at $\sqrt{s} = 8$~\TeV, by ATLAS using the $\tau\nu$ decay mode~\cite{taunupaper} and by CMS using final states originating from both the $\tau\nu$ and $tb$ decay modes~\cite{PAS-HIG-14-023}.
CMS set an upper limit of 2.0--0.13~pb on the production cross section times branching fraction for $H^+ \to tb$ in the mass range 180--600~\GeV.
Vector-boson-fusion $H^{+}$ production was also searched for by ATLAS using the $WZ$ final state~\cite{WZpaper}. No evidence for a charged Higgs boson was found in any of these searches.

This paper describes searches for charged Higgs bosons decaying into $tb$.
In the $H^+$ mass range of 200--600~\GeV, the production mode in association with a top quark is studied.
The 5FS process is generated.
Cross sections averaging 4FS and 5FS are used for model-dependent predictions. The search is based on
selecting two top quarks, with their decays producing one charged lepton (electron or muon), and at
least one additional jet containing a $b$-flavoured hadron. In the complementary $s$-channel production mode,
\Hplus\ masses between 0.4 and 2.0~\TeV\ are explored in a final state containing one charged lepton
and jets (referred to as lepton+jets in the following), while the all-hadronic final state is used for very
high $H^+$ masses, 1.5 to 3.0~\TeV, with a jet substructure technique to reconstruct the
top-quark decay products in one single large-radius jet. The two $s$-channel analyses are
reinterpretations of recent searches for $\Wprime \to tb$ in ATLAS~\cite{Aad:2014xea,Aad:2014xra}.
Based on dedicated simulations of the $H^+ \to tb$ signal and a reinterpretation of the data,
upper limits are derived for the $s$-channel production of a charged scalar particle decaying to $tb$.

The paper is organised as follows. Section~\ref{sec:Samples} describes briefly the ATLAS
detector, then summarises the data and the samples of simulated events used for the
analyses. Section~\ref{sec:PhysicsObjects} describes the reconstruction of objects in ATLAS.
Section~\ref{sec:Selection} presents the event selection and analysis strategy of
the search for $H^+ \to tb$ produced in association with a top quark. Systematic uncertainties
are also discussed, before exclusion limits in terms of cross section times branching
fraction are presented, together with their interpretation in benchmark scenarios of
the Minimal Supersymmetric Standard Model (MSSM)~\cite{mssm1,mssm2,mssm3,mssm4,mssm5}.
The reinterpretations of $\Wprime \to tb$ analyses as searches for the production of
$\Hplus \to tb$ in the $s$-channel, including a discussion of the $H^+$ signal shapes and
uncertainties, are presented in section~\ref{sec:schannel}. Finally, a summary is
given in section~\ref{sec:Conclusion}.

\section{Data and simulated events}
\label{sec:Samples}

\subsection{ATLAS detector and data sample}
\label{sec:ATLAS_detector_and_data}

The ATLAS detector~\cite{Aad:2008zzm} consists of an inner tracking
system with coverage in pseudorapidity\footnote{
ATLAS uses a right-handed coordinate system with
its origin at the nominal interaction point (IP) in the centre
of the detector and the $z$-axis along the beam pipe. The $x$-axis
points from the IP to the centre of the LHC ring, and the $y$-axis
points upwards. Cylindrical coordinates $(r,\phi)$ are used in the
transverse plane, $\phi$ being the azimuthal angle around the $z$-axis.
The pseudorapidity is defined in terms of the polar angle $\theta$ as
$\eta=-\ln\tan(\theta/2)$.}
up to $|\eta|=2.5$, surrounded by a thin 2~T superconducting solenoid,
a calorimeter system extending up to $|\eta|=4.9$ and a muon spectrometer
extending up to $|\eta|=2.7$ that measures the deflection of muon
tracks in the field of three superconducting toroid magnets. A
three-level trigger system is used to select events of interest.
The first-level trigger (L1) is implemented in hardware, using a
subset of detector information to reduce the event rate to no more
than 75~kHz. This is followed by two software-based trigger levels
(L2 and EF), which together further reduce the event rate to less than 400~Hz.

Stringent data-quality requirements are applied,
resulting in an integrated luminosity of $\unit[20.3]{\ifb}$
for the 2012 data-taking period. The integrated luminosity
has an uncertainty of 2.8\%, measured following the methodology
described in ref.~\cite{lumi}. Events are required to have
a primary vertex with at least five associated tracks, each with
a transverse momentum $\pt$ greater than 400~\MeV. If an event has
more than one reconstructed vertex satisfying these criteria, the
primary vertex is defined as the reconstructed vertex with the largest sum of squared
track transverse momenta.

\subsection{Background and signal modelling}
\label{sec:background_and_signal_modelling}

The background processes for the searches in this paper include SM pair production of top quarks
(with additional jets, or in association with a vector boson $V=W,Z$ or the SM Higgs boson), as well as
the production of single-top-quark, $W$+jets, $Z/\gamma^{\ast}$+jets, diboson ($WW/WZ/ZZ$) and
multi-jet events. The dominant background is the production of $t\bar{t}$ pairs with additional jets in the final state.

In the analyses with an electron or a muon in the final state, all backgrounds are taken from
simulation, except for the multi-jet events. These mostly contribute via the presence of a
non-prompt electron or muon, e.g.\ from a semileptonic $b$- or $c$-flavoured hadron decay, or
through the misidentification of a jet. The normalisation of the multi-jet events and the shape of the relevant distributions are
determined with a data-driven technique known as the matrix method~\cite{MM}. In the search
for $H^+ \to tb$ in the $s$-channel production mode with an all-hadronic final state, all
backgrounds are estimated using a data-driven method based on a combined fit to the data
under the SM background plus \Hplus\ signal hypothesis.

The modelling of $\ttbar$ events is performed with
{\sc Powheg-Box} v2.0~\cite{powheg,powbox2}, using the
CT10~\cite{ct10one,ct10two} parton distribution function
(PDF) set. It is interfaced to {\sc Pythia} v6.425~\cite{Pythia6},
with the Perugia P2011C~\cite{p2011c} set of tuned parameters (tune) for the underlying event.
The $t\bar{t}$ cross section at 8~\TeV\ is
$\sigma_{t\bar{t}}= 253^{+13}_{-15}$~pb for a top-quark mass of
172.5~\GeV. It is calculated at next-to-next-to-leading order
(NNLO) in QCD including resummation of next-to-next-to-leading
logarithmic (NNLL) soft gluon terms with
{\sc Top}++ v2.0~\cite{Cacciari:2011hy,Beneke,Barnreuther,Czakon1,Czakon2,Czakon3,Czakon:2011xx}.

In the search for $H^+$ production in association with a top quark, simulated $\ttbar$ events are classified according to their flavour content at parton level, using the same methodology as in ref.~\cite{tth}. Events are labelled as $t\bar{t}$+$b\bar{b}$ if they contain at least one particle jet that is matched to a $b$-flavoured hadron not originating from the decay of the $t\bar{t}$ system. Events where at least one particle jet is matched to a $c$-flavoured hadron, and not already labelled as $t\bar{t}$+$b\bar{b}$, are labelled as $t\bar{t}$+$c\bar{c}$.
Events labelled as either $t\bar{t}$+$b\bar{b}$ or $t\bar{t}$+$c\bar{c}$ are generically referred to as $t\bar{t}$+heavy-flavour (HF) events.
The remaining events, including those with no additional jets, are labelled as $\ttbar$+light-flavour (LF).
In the following, a sequential reweighting is applied at the generator level for
all $t\bar{t}$+LF and $t\bar{t}$+$c\bar{c}$ events produced with {\sc Powheg+Pythia}.
Two correction factors are used, based on the values of the transverse
momenta of the top quark and the $t\bar{t}$ system, taking the correlation
between these two parameters into account. This reweighting procedure was
originally implemented in order to match simulation to data in the
measurement of top-quark-pair differential cross sections at
$\rts = 7~\mbox{\TeV}$~\cite{top7tev}. It was verified that this
procedure is also reasonable at $\rts = 8~\mbox{\TeV}$. The $t\bar{t}$+$b\bar{b}$ component is reweighted to match the NLO theory calculation provided within {\sc Sherpa} with the {\sc OpenLoops} framework~\cite{sherpa1,sherpa2}.
For this reweighting, the same settings as in ref.~\cite{tth} are used in this paper.
The reweighting is performed at the generator level using several kinematic variables such as the transverse momenta of the top quark, the $t\bar{t}$ system and the dijet system not coming from the top-quark decay, as well as the distance\footnote{$\Delta R=\sqrt{(\Delta \eta)^2 + (\Delta \phi)^2}$, where $\Delta \eta$ is the difference in pseudorapidity of the two objects in question, and $\Delta \phi$ is the difference between their azimuthal angles.} $\Delta R_{jj}$ between these two jets.
For systematic studies, an alternative $t\bar{t}$+jets sample is generated with the {\sc Madgraph5}~v1.5.11 LO generator~\cite{madgraph}, using the CT10 PDF set and interfaced to {\sc Pythia}~v6.425 for parton shower and fragmentation.

Samples of $t\bar{t}V$ events are generated using {\sc Madgraph5}~v1.3.33, with the CTEQ6L1~\cite{cteq} PDF, interfaced to {\sc Pythia}~v6.425 for the showering and hadronisation, 
with the AUET2B underlying-event tune~\cite{PUB-2011-009}. They are normalised
to the next-to-leading-order (NLO) cross section~\cite{ttV1,ttV2}.

Single-top-quark production in the $s$- and $Wt$-channels are
simulated with {\sc Powheg-Box}~v2.0, using the CT10 PDF, interfaced to
{\sc Pythia}~v6.425 with the underlying-event tune P2011C. The same
procedure is used for the single-top-quark production in the $t$-channel,
except in the search for $q\qprime \to H^+ \to tb$ in the lepton+jets final
state, where the leading-order (LO) generator {\sc AcerMC}~v3.8~\cite{KER-0401}
with the CTEQ6L1 PDF, interfaced to {\sc Pythia}~v6.425 with the
underlying-event tune P2011C, is used instead. Overlaps between
the \ttbar\ and $Wt$ final states are handled using inclusive diagram removal~\cite{Frixione2008}.
The single-top-quark samples are normalised to the approximate NNLO
theoretical cross sections~\cite{stopxs1,stopxs2,stopxs3} using the
MSTW2008 NNLO~\cite{Martin:2009iq,Martin:2009bu,Martin:2010db} PDF.

Samples of $W/Z$+jets events are generated using the
{\sc Alpgen}~v2.14~\cite{alpgen} generator, with the CTEQ6L1
PDF, interfaced to {\sc Pythia}~v6.425 with the underlying-event
tune P2011C. The $W$+jets events are generated with up to five
additional partons, separately for the $W$+LF, $Wb\bar{b}$+jets,
$Wc\bar{c}$+jets and $Wc$+jets processes. Similarly, the $Z$+jets background
is generated with up to five additional partons separated in different
flavours.
The samples of $W/Z$+jets events are normalised
to the inclusive NNLO theoretical cross sections~\cite{vjetsxs}.
Finally, the $W/Z$+jets events are reweighted to account for
differences in the $W/Z$ \pt\ spectrum between data and
simulation~\cite{Zjetsxsec}.

In the searches for $H^+ \to tb$ with a lepton+jets final state, diboson events
are generated with the requirement of having at least one boson decaying leptonically.
{\sc Alpgen}~v2.14 is used, with the CTEQ6L1 PDF, and it is interfaced to
{\sc Herwig}~v6.520~\cite{COR-0001} for showering and hadronisation, together
with {\sc Jimmy}~v4.31~\cite{Butterworth:1996zw} for the underlying-event,
using the AUET2 tune~\cite{PUB-2011-008}. The diboson backgrounds are normalised
to the production cross sections calculated at NLO~\cite{dibosonxs}.

The production of the SM Higgs boson in association with a top-quark pair ($\ttbar H$) is
modelled using NLO matrix elements obtained from the {\sc HELAC-Oneloop} package~\cite{Helac}.
{\sc Powheg-Box} is used as an interface to shower simulation programs. The samples created
using this approach are referred to as {\sc PowHel} samples~\cite{ttH_NLO}. They are inclusive
in Higgs boson decays and are produced for a Higgs boson mass of 125~\GeV, using the CT10 PDF,
and interfaced to {\sc Pythia}~v8.1~\cite{Sjostrand:2007gs} with the AU2 underlying-event
tune~\cite{au2}. As in the generation of $\ttbar$ background events, the top-quark mass
is set to 172.5~\GeV. The $t\bar{t}H$ cross section and the decay branching fractions of the
Higgs boson are taken from the (N)NLO theoretical calculations collected in ref.~\cite{Dittmaier:2011ti}.

In the search for $H^{+}$ produced in association with a top quark, signal samples are generated
with {\sc Powheg-Box}, using the CT10 PDF, interfaced to {\sc Pythia}~v8.1 with the AU2
underlying-event tune. For the $m_{H^+}$ range of 200--300~\GeV, the samples are produced in
steps of 25~\GeV, then in intervals of 50~\GeV\ up to 600~\GeV. The samples are generated at NLO
using the 5FS and with a zero width for $\Hp$.

In the search for $H^+$ in the $s$-channel, signal events are generated using {\sc Madgraph5}~v1.5.12, with the CTEQ6L1 PDF, interfaced to {\sc Pythia}~v8.1 with the AU2 underlying-event tune, for both the lepton+jets and all-hadronic final states. In the former (latter) case, samples are produced in $m_{H^+}$ steps of 200 (250)~\GeV, between 0.4 and 2.0~\TeV\ (1.5 and 3.0~\TeV).
A narrow-width approximation is used for both final states.
This is justified as the experimental resolution is much larger than the $H^+$ natural width.

In all background simulations, {\sc Tauola}~v1.20~\cite{Was:2004dg} is used for
the $\tau$ decays and {\sc Photos}~v2.15~\cite{Barberio:1990ms} is employed
for photon radiation from charged leptons. For the signal simulations, {\sc Photos++}~v3.51~\cite{photospp} is used. All signal
and background events are overlaid with additional minimum-bias events
generated using {\sc Pythia}~v8.1 with the MSTW2008 LO PDF and the AUET2
underlying-event tune, in order to simulate the effect of multiple $pp$
collisions per bunch crossing (pile-up).
Finally, all background samples and all-hadronic signal samples are processed
through a simulation~\cite{simulation} of the detector geometry and response using
{\sc Geant4}~\cite{AGO-0301}. The signal samples with leptons in the final state
are passed through a fast simulation of the calorimeter response~\cite{fastsim}.
All samples from simulation are processed through the same reconstruction software as the data.

\section{Object reconstruction and identification}
\label{sec:PhysicsObjects}

The main objects used for the searches reported in this
paper are electrons, muons, jets (possibly identified as originating from
$b$-quarks), and missing transverse momentum. A brief summary of the main
reconstruction and identification criteria used for each of
these objects is given below.

Electron candidates~\cite{ElectronPerformance} are reconstructed from
energy deposits (clusters) in the electromagnetic calorimeter which
are associated with a reconstructed track in the inner detector system. Their
transverse energy, $E_{\text{T}}=E_{\text{clus}}/{\rm cosh}(\eta_{\text{track}})$,
is computed using the electromagnetic cluster energy $E_{\text{clus}}$ and the
direction of the electron track $\eta_{\text{track}}$, and is required to exceed
25~\GeV. The pseudorapidity range for the electromagnetic cluster covers the fiducial
volume of the detector, $|\eta|<2.47$ (the transition region between the barrel
and end-cap calorimeters, $1.37<|\eta|<1.52$, is excluded). The longitudinal
impact parameter $|z_{0}|$ of the electron track relative to the primary
vertex must be smaller than $\unit[2]{mm}$. In order to reduce the
contamination from misidentified hadrons, electrons from heavy-flavour
decays and photon conversions, the electron candidates are also required
to satisfy $\ET$- and $\eta$-dependent calorimeter (and tracker) isolation
requirements imposed in a cone with a fixed size  $\Delta R = 0.2~(0.3)$
around the electron position.

Muon candidates are reconstructed from track segments in the
muon spectrometer, and matched with tracks found in the inner
detector system~\cite{muons}. The final muon candidates are refitted
using the complete track information from both detector systems,
and they are required to satisfy $\pt > \unit[25]{\GeV}$, $|\eta|<2.5$ and
$|z_{0}| < \unit[2]{mm}$.
Furthermore, muons must fulfil a $\pt$-dependent track-based isolation requirement
that has good performance under high pile-up conditions and/or when the muon is
close to a jet. For that purpose, the scalar sum of the track $\pt$ in a cone of
a variable size, defined by $\Delta R = \unit[10]{\GeV}/\pt$, around the muon
position (while excluding the muon track itself) must be less than $5\%$ of
the muon transverse momentum.

Jets are reconstructed from topological energy clusters~\cite{lampl} in the
calorimeters, using the anti-\kt\ algorithm~\cite{Cacciari:2008gp, Cacciari:2005hq}.
Two radius parameters are used, $R=0.4$ ('small-radius jets') or $R=1.0$
('large-radius jets').
The large-radius jets are only used when reconstructing high-$\pt$
top quarks as single objects in the search for $H^+ \to tb$ produced in the $s$-channel
and decaying into an all-hadronic final state, as described below.
When no jet type is specified, small-radius jets are implied.
Small- and large-radius jets are calibrated using energy- and $\eta$-dependent correction
factors derived from simulation and with residual corrections from in situ measurements~\cite{JES14060076}.
Only small-radius jets that have $\pt > \unit[25]{\GeV}$ and $|\eta|<2.5$ are considered
in this paper.
Jets originating from pile-up interactions are suppressed by requiring that at least 50\% of the
scalar sum of the \pt\ of the associated tracks is due to tracks originating from the primary vertex~\cite{JVFref}.
This is referred to as the jet vertex fraction (JVF) and is only applied to jets with
$\pt < \unit[50]{\GeV}$ and $|\eta |<2.4$.

Jets are identified as originating from the hadronisation of a $b$-quark
($b$-tagged) via an algorithm that
uses multivariate techniques to combine information from the impact
parameters of displaced tracks with topological properties of secondary
and tertiary decay vertices reconstructed within the jet~\cite{bib:AdvancedTaggers}.
The nominal working point used here is chosen to correspond to a 70\% efficiency
to tag a $b$-quark jet, with a light-jet mistag rate of 1\% and a $c$-jet
mistag rate of 20\%, as determined with $b$-tagged jets with
$\pt > \unit[20]{\GeV}$ and $|\eta|<2.5$ in simulated $\ttbar$
events. The tagging efficiencies from simulation are corrected
based on the results of flavour-tagging calibrations performed
with the data~\cite{btag-data}.

\label{sec:toptagging}
In the search for $H^+ \to tb$ produced in the $s$-channel and decaying into
an all-hadronic final state (section~\ref{sec:SchanAllHad}), hadronically
decaying high-$\pt$ top quarks are reconstructed as single objects through
'top-tagging'. Large-radius jets
are used as input to the top-tagger. In order to minimise the effects of
pile-up~\cite{ATLAS-CONF-2012-066}, the large-radius jets
are trimmed~\cite{Krohn:2009th}. The trimming is performed by reclustering
the large-radius jet using the inclusive \kt\ algorithm~\cite{Ellis:1993tq}
with a jet radius parameter
$R = 0.3$, and by removing soft subjets with a \pt\ smaller than 5\% of the
original jet $\pt$. Trimmed large-radius jets are required to have
$\pt > \unit[350]{\GeV}$ and $|\eta|<2.0$.
Large-radius jets are top-tagged if they have a substructure compatible with a
three-prong decay. The top-tagger used in the search of section~\ref{sec:SchanAllHad}
was developed for the search for $\Wprime \rightarrow tb$ in ATLAS~\cite{Aad:2014xra}.
It uses the \kt\ splitting scale~\cite{Ellis:1993tq} $\sqrt{d_{12}}$ and the
$N$-subjettiness~\cite{Thaler:2010tr,Thaler:2011gf} variables $\tau_{21}$ and $\tau_{32}$.
The \kt\ algorithm clusters the hardest objects last, which means that a two-body decay
(such as $t \rightarrow b W$) typically gets a larger value of $\sqrt{d_{12}}$ than light jets.
The $\tau_{ij}$ distribution peaks closer to 0 for $i$-subjet-like jets and closer to 1 for
$j$-subjet-like jets. The top-tagged jet is required to pass the cuts $\sqrt{d_{12}}>40$~\gev,
$\tau_{32}< 0.65$, and $0.4 <\tau_{21}<0.9$, as in the search for $\Wprime \to tb$~\cite{Aad:2014xra}.

When several selected objects overlap geometrically, the following procedures are applied.
In the searches with a lepton+jets final state, muons are rejected if found to be $\Delta R < 0.4$
from any jet with nominal $\pt$, $\eta$ and JVF selections. In order to avoid double-counting of
electrons as jets, the closest jet to an electron is then removed if lying $\Delta R < 0.2$
from an electron. Finally, electrons are rejected if found to be $\Delta R < 0.4$ from any remaining jet with
nominal $\pt$, $\eta$ and JVF selections. In the search for $s$-channel production of $H^+ \to tb$
in the all-hadronic final state, large-radius jets are required to be separated by $\Delta R > 2.0$
from the small-radius $b$-tagged jets used to reconstruct the invariant mass of \Hplus\ candidates.
Events with electrons (muons) fulfilling $\ET > \unit[30]{\GeV}$ ($\pt > \unit[30]{\GeV}$) are vetoed
in this particular search channel.

The magnitude $\met$ of the missing transverse momentum is reconstructed from the negative vector
sum of transverse momenta of reconstructed objects, as well as from unmatched topological clusters
and tracks (collected in a so-called soft term).
The $\met$ is further refined by using object-level corrections for the identified electrons, muons and jets,
and the effects of pile-up in the soft term are mitigated~\cite{etmiss}.

\section{Search for a charged Higgs boson in association with a top quark}
\label{sec:Selection}

\subsection{Event selection and categorisation}
\label{sec:SelectionCutbased}

In this section, the search for a charged Higgs boson produced in association with a top quark, $gb\to tH^{+}$ with $H^+ \to tb$, is described. In the events selected for this analysis, the top quarks both decay via $t\to Wb$, where one $W$ boson decays hadronically and the other decays into an electron or a muon, either directly or through a $\tau$-lepton decay, and the corresponding neutrino(s). The signal event signature is therefore characterised by the presence of exactly one high-$\pt$ charged lepton (electron or muon) and five or more jets, at least three of them being $b$-tagged.

Events collected using either an isolated or non-isolated single-lepton trigger are considered. Isolated triggers have a threshold of 24~\GeV\ on $p_\mathrm{T}$ for muons and on $E_\mathrm{T}$ for electrons, while non-isolated triggers have higher thresholds at 36~\GeV\ (muons) and 60~\GeV\ (electrons). The isolated triggers have a loss of efficiency at high $p_\mathrm{T}$ or $E_\mathrm{T}$, which is recovered by the triggers with higher thresholds.
Events accepted by the trigger are then required to have exactly one identified electron or muon, and at least four jets, of which at least two must be identified as $b$-tagged jets. The selected lepton is required to match, with $\Delta R < 0.15$, a lepton reconstructed by the trigger.

At this stage, the samples contain mostly $\ttbar$ events. The selected events are further categorised into different regions, depending on the number of jets and $b$-tagged jets. The categories are inclusive in the lepton flavour. In the following, a given category with $m$ jets, of which $n$ are $b$-tagged, is referred to as $m$j($n$b). A total of five independent categories are considered: four control regions (CR) with little sensitivity to signal, 4j(2b), 5j(2b), $\geq$6j(2b), 4j($\geq$3b), and one signal-rich region (SR), $\geq$5j($\geq$3b). The CR are used to control the backgrounds and to constrain systematic uncertainties (section~\ref{sec:ResolvedSystematics}). For each category, the expected event yields of all processes and the number of events observed in the data are given in table~\ref{tab:tH_prefityields}. The dominant background process in every category is $t\bar{t}$+LF. In the signal-rich region, contributions from $t\bar{t}$+HF are also sizeable. In all categories except $\geq$6j(2b), the data exceed the SM prediction, but they are consistent within the large uncertainties on the background. In table~\ref{tab:tH_benchmarksignal}, the expected amount of signal is listed for a few points of the $m_h^\mathrm{mod-}$ benchmark scenario of the MSSM~\cite{MSSM}. The theoretical predictions are taken from refs.~\cite{xsec1,xsec2,xsec3,xsec4}.

\begin{table}[h!]
\begin{center}
\resizebox{\textwidth}{!}{%
\begin{tabular}{c
S[table-format=6.1, table-number-alignment=right]@{\,}@{$\pm$}@{\,}S[table-format=5.1, table-number-alignment=right]
S[table-format=5.1, table-number-alignment=right]@{\,}@{$\pm$}@{\,}S[table-format=4.1, table-number-alignment=right]
S[table-format=5.1, table-number-alignment=right]@{\,}@{$\pm$}@{\,}S[table-format=4.1, table-number-alignment=right]
S[table-format=4.1, table-number-alignment=right]@{\,}@{$\pm$}@{\,}S[table-format=4.1, table-number-alignment=right]
S[table-format=5.1, table-number-alignment=right]@{\,}@{$\pm$}@{\,}S[table-format=4.1, table-number-alignment=right]}
\toprule
Process & \multicolumn{2}{c}{4j(2b)} & \multicolumn{2}{c}{5j(2b)} & \multicolumn{2}{c}{$\geq$6j(2b)} & \multicolumn{2}{c}{4j($\geq$3b)} & \multicolumn{2}{c}{$\geq$5j($\geq$3b)}\\
\hline
$t\bar{t}$+LF & 80300 & 9900 & 38700 & 7400 & 19300 & 5300 & 6300 & 1000 & 5600  & 1600  \\
$t\bar{t}$+$c\bar{c}$ & 5200 & 2900 & 4500 & 2600 & 3800 & 2300 & 740 & 410 & 1800 & 1000 \\
$t\bar{t}$+$b\bar{b}$ & 1720 & 940  & 1550 & 830  & 1390 & 820  & 660 & 370 & 2300 & 1200 \\
$t\bar{t}H$   & 33.7  & 4.6  & 44.6  & 5.4  & 68.9 & 9.1  & 15.5 & 2.5 & 87   & 11 \\
$t\bar{t}V$   & 128   & 40   & 151   & 47   & 189  & 59   & 17.6 & 5.7 & 85   & 27 \\
Single-top    & 5020  & 770  & 1970  & 420  & 880  & 270  & 360  & 83  & 330  & 110 \\
$W$+jets        & 3400  & 1700 & 1270  & 720  & 640  & 400  & 190  & 100 & 170  & 100 \\
$Z$+jets        & 1330  & 670  & 400   & 220  & 150  & 95   & 53   & 31  & 49   & 39 \\
$VV$            & 232   & 69   & 108   & 41   & 52   & 25   & 10.7 & 3.6 & 13.7 & 6.0 \\
Multi-jets    & 2160  & 870  & 670   & 260  & 330  & 150  & 160  & 67  & 150  & 100 \\
\hline
Total bkg & 100000 & 11000 & 49300 & 8600 & 27100 & 6600 & 8500 & 1300 & 10600 & 2500  \\
\hline
Data & \multicolumn{2}{l}{\num{102462}} & \multicolumn{2}{l}{\num{51421}} & \multicolumn{2}{l}{\num{26948}} & \multicolumn{2}{l}{\num{9102}} & \multicolumn{2}{l}{\num{11945}}  \\
\bottomrule
\end{tabular}}
\caption{\label{tab:tH_prefityields} Expected event yields of the SM background processes and observed data in the five categories. The first four columns show the event yields in the CR, the last column shows the event yields in the SR. The uncertainties include statistical and systematic components (systematic uncertainties are discussed in section~\ref{sec:ResolvedSystematics}).}
\end{center}
\end{table}

\begin{table}[h!]
\begin{center}
\resizebox{\textwidth}{!}{%
\begin{tabular}{cc
S[table-format=4.1, table-number-alignment=right]@{\,}@{$\pm$}@{\,}S[table-format=3.1, table-number-alignment=right]
S[table-format=4.1, table-number-alignment=right]@{\,}@{$\pm$}@{\,}S[table-format=3.1, table-number-alignment=right]
S[table-format=4.1, table-number-alignment=right]@{\,}@{$\pm$}@{\,}S[table-format=3.1, table-number-alignment=right]
S[table-format=3.1, table-number-alignment=right]@{\,}@{$\pm$}@{\,}S[table-format=3.1, table-number-alignment=right]
S[table-format=4.1, table-number-alignment=right]@{\,}@{$\pm$}@{\,}S[table-format=3.1, table-number-alignment=right]}
\toprule
$m_{H^+}$ [\GeV] & $\tan\beta$ & \multicolumn{2}{c}{4j(2b)} & \multicolumn{2}{c}{5j(2b)} & \multicolumn{2}{c}{$\geq$6j(2b)} & \multicolumn{2}{c}{4j($\geq$3b)} & \multicolumn{2}{c}{$\geq$5j($\geq$3b)} \\
\hline
\multirow{3}{*}{200} & 0.5 & 2580 & 420 & 1670 & 190 & 1050 & 300 & 730  & 190 & 1750 & 200 \\
                     & 0.7 & 1290 & 210 & 834  & 93  & 520  & 150 & 366  & 95  & 880  & 100 \\
                     & 0.9 & 760  & 120 & 493  & 55  & 309  & 88  & 216  & 56  & 518  & 59  \\
\hline
\multirow{3}{*}{400} & 0.5 & 397  & 69  & 406  & 44  & 390  & 100 & 211  & 56  & 756  & 76  \\
                     & 0.7 & 200  & 35  & 204  & 22  & 197  & 51  & 106  & 28  & 380  & 38  \\
                     & 0.9 & 119  & 21  & 121  & 13  & 117  & 31  & 63   & 17  & 226  & 23  \\
\hline
\multirow{3}{*}{600} & 0.5 & 71   & 14  & 85   & 12  & 107  & 29  & 36   & 11  & 183  & 23  \\
                     & 0.7 & 34.7 & 6.9 & 41.5 & 5.6 & 52   & 14  & 17.4 & 5.3 & 89   & 11  \\
                     & 0.9 & 19.8 & 3.9 & 23.7 & 3.2 & 29.8 & 8.1 & 10.0 & 3.0 & 50.9 & 6.5 \\
\bottomrule
\end{tabular} }
\caption{Number of expected signal events in the five categories for a few representative points of the $m_h^\mathrm{mod-}$ scenario of the MSSM. The last column shows the event yields in the SR. The expected uncertainties contain statistical and systematic components (systematic uncertainties are discussed in section~\ref{sec:ResolvedSystematics}). Uncertainties on the cross sections and branching fractions for the $m_h^\mathrm{mod-}$ scenario are not included. \label{tab:tH_benchmarksignal}}
\end{center}
\end{table}

\subsection{Analysis strategy}
\label{sec:ResolvedStrategy}

In order to separate the $H^+$ signal from the SM background, and to constrain the large uncertainties on the background, different discriminants are used depending on the event category, and are then combined in a binned maximum-likelihood fit. In the four CR, the discriminating variable is the scalar sum of the $\pt$ of the selected jets ($H_\mathrm{T}^\mathrm{had}$) and in the SR, the output of a boosted decision tree (BDT) is used. The Toolkit for Multivariate Data Analysis ({\sc TMVA})~\cite{tmva} is used for the training and evaluation of the BDT responses. The BDT is trained to specifically discriminate the $H^{+}$ signal from the $t\bar{t}$+$b\bar{b}$ background process. This method reduces correlations and anti-correlations between the signal normalisation and the parameters connected to the dominant systematic uncertainties, in particular for $H^{+}$ masses below 350~\GeV, where those correlations are sizeable. The largest correlation at low mass is that between the $t\bar{t}$+$b\bar{b}$ cross section and the signal normalisation, which is $-$50\% at 200~\GeV. Consequently, this specific BDT is more sensitive than a BDT trained against the sum of all backgrounds when uncertainties are included. The variables entering the BDT training are:
\begin{itemize}
\item the scalar sum of the $\pt$ of all selected jets ($H_{\mathrm{T}}^{\mathrm{had}}$),
\item the $p_\mathrm{T}$ of the leading jet,
\item the invariant mass of the two $b$-tagged jets that are closest in $\Delta R$,
\item the second Fox--Wolfram moment~\cite{fox}, calculated from the selected jets,
\item the average $\Delta R$ between all pairs of $b$-tagged jets in the event.
\end{itemize}
Many other kinematic and event shape variables were tested before this set of variables was selected. The variables listed above provide the best separation between signal and background across all mass hypotheses. The BDT training is performed independently for each $H^+$ mass hypothesis, and only for events in the SR. The BDT input variables were validated in the CR by comparing their distributions in the data and simulation, and they were further validated by evaluating the BDT responses in the four CR for every mass point. The data and expected SM backgrounds were found to be compatible at all times. The statistical analysis was performed after the selection and the BDT training were finalised.

\begin{figure}[h!]
  \begin{center}
  \subfloat[]{ \includegraphics[width=0.4\textwidth]{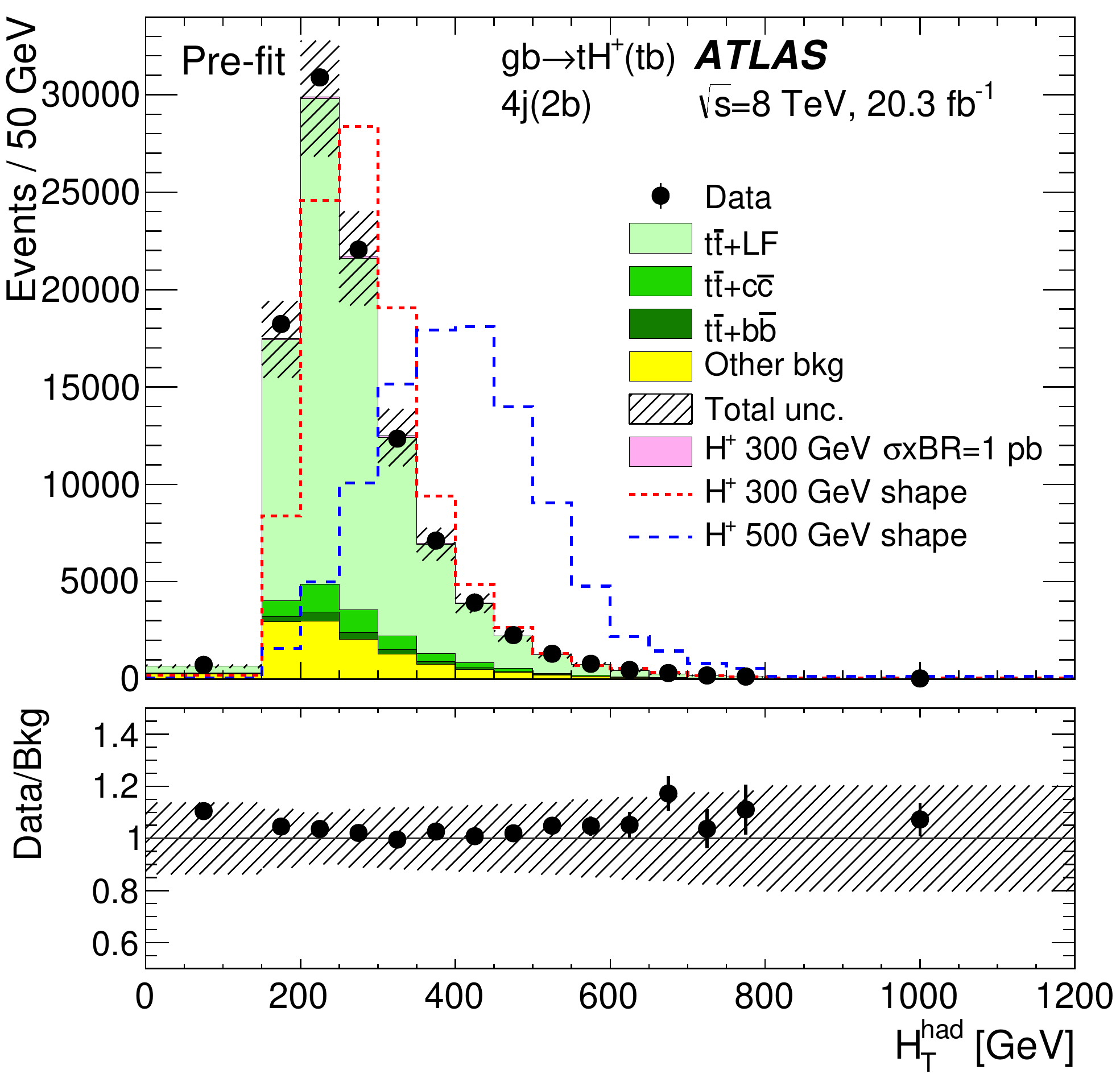} }
  \subfloat[]{ \includegraphics[width=0.4\textwidth]{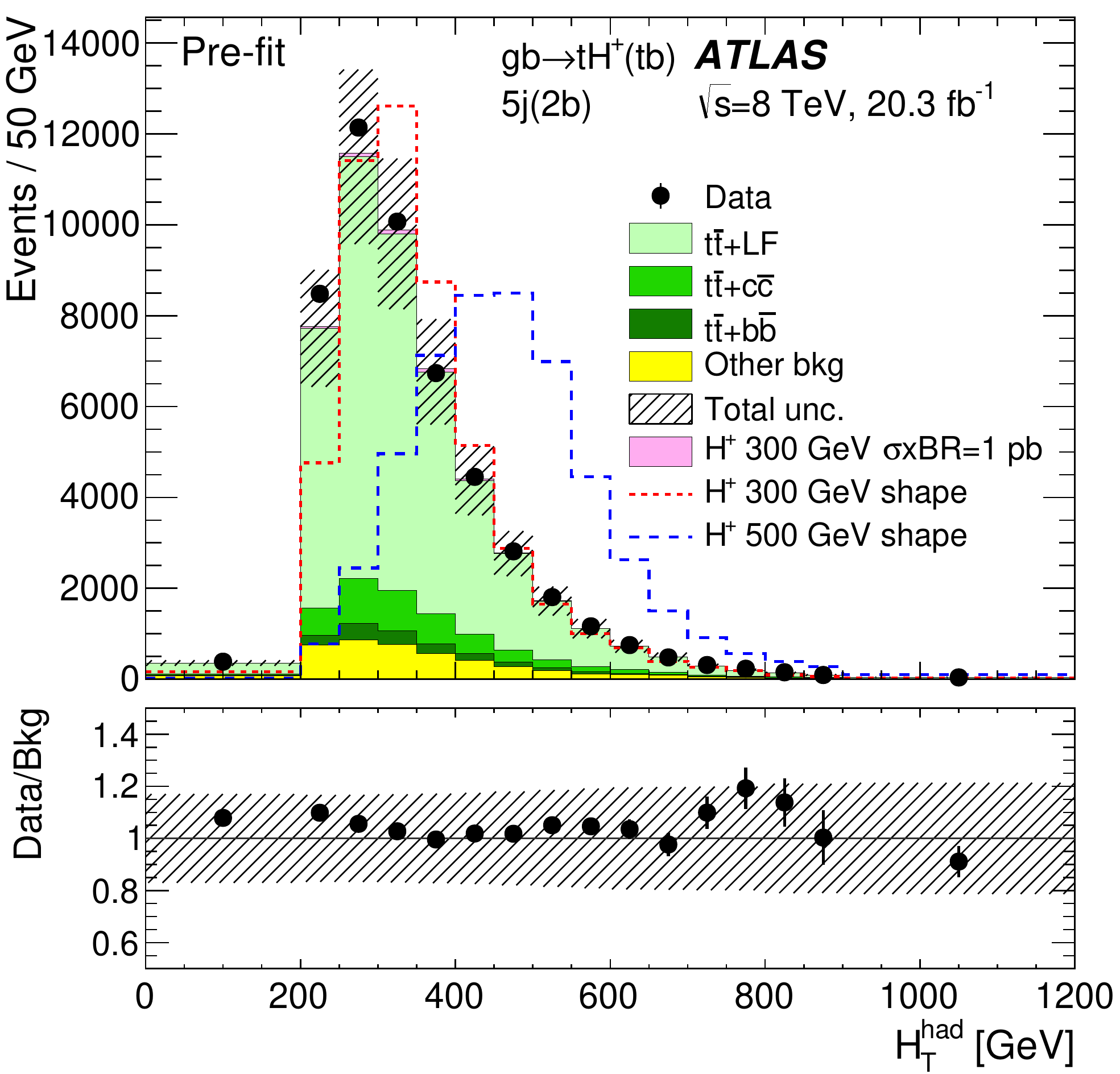} }\\
  \subfloat[]{ \includegraphics[width=0.4\textwidth]{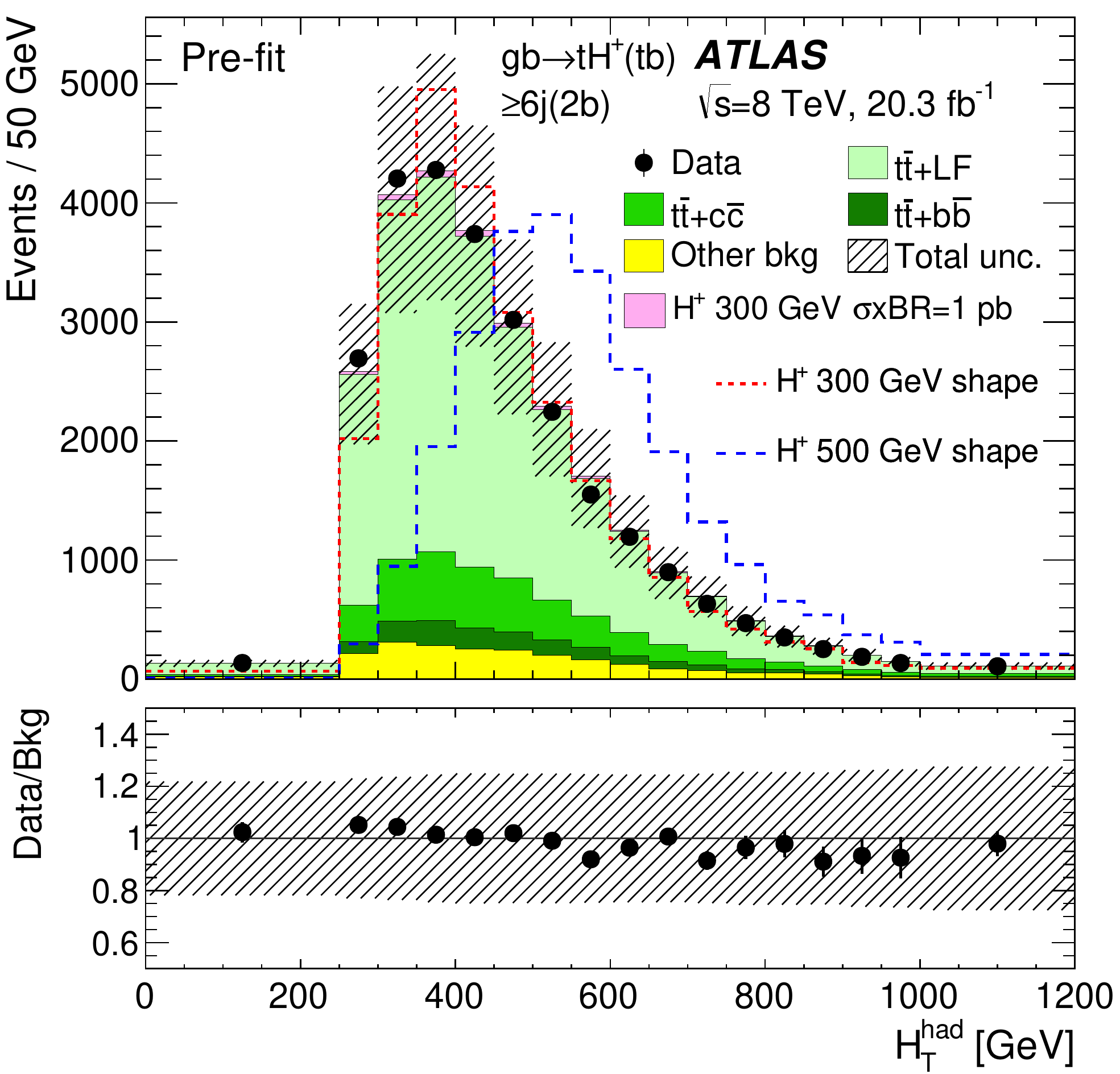} }
  \subfloat[]{ \includegraphics[width=0.4\textwidth]{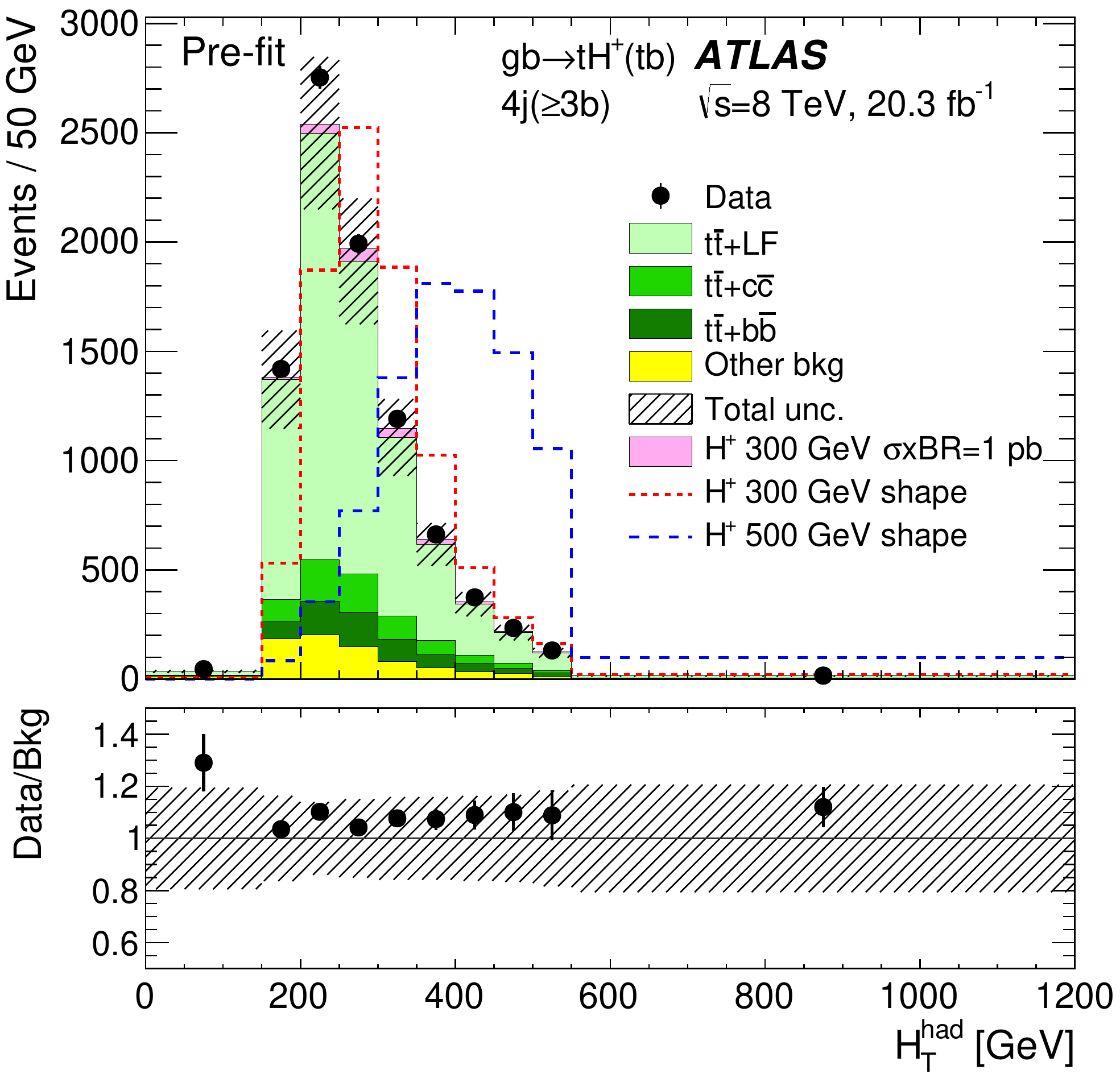} }
  \end{center}
  \caption{Pre-fit distributions of the scalar sum of the $p_\mathrm{T}$ of all selected jets, $H_\mathrm{T}^\mathrm{had}$, for the four control regions:
(a) 4j(2b), (b) 5j(2b), (c) $\geq$6j(2b), (d) 4j($\geq$3b). Each background process is normalised
according to its cross section. A signal with $m_{H^{+}}=300$~\GeV, normalised to a production cross
section times branching fraction for $H^+ \to tb$ ($\sigma \times \mbox{BR}$) of 1~pb, is shown in
pink, stacked on top of the background. Two signal shapes are shown superimposed as dashed lines normalised to the data. The last bin includes the
overflow. The hatched bands show the pre-fit uncertainties, which are dominated by systematic
uncertainties (discussed in section~\ref{sec:ResolvedSystematics}). The lower panels display the
ratio of the data to the total predicted background.  \label{fig:tH_hthadprefit}}
\end{figure}

\begin{figure}[h!]
\centering
\subfloat[]{ \includegraphics[width=0.49\textwidth]{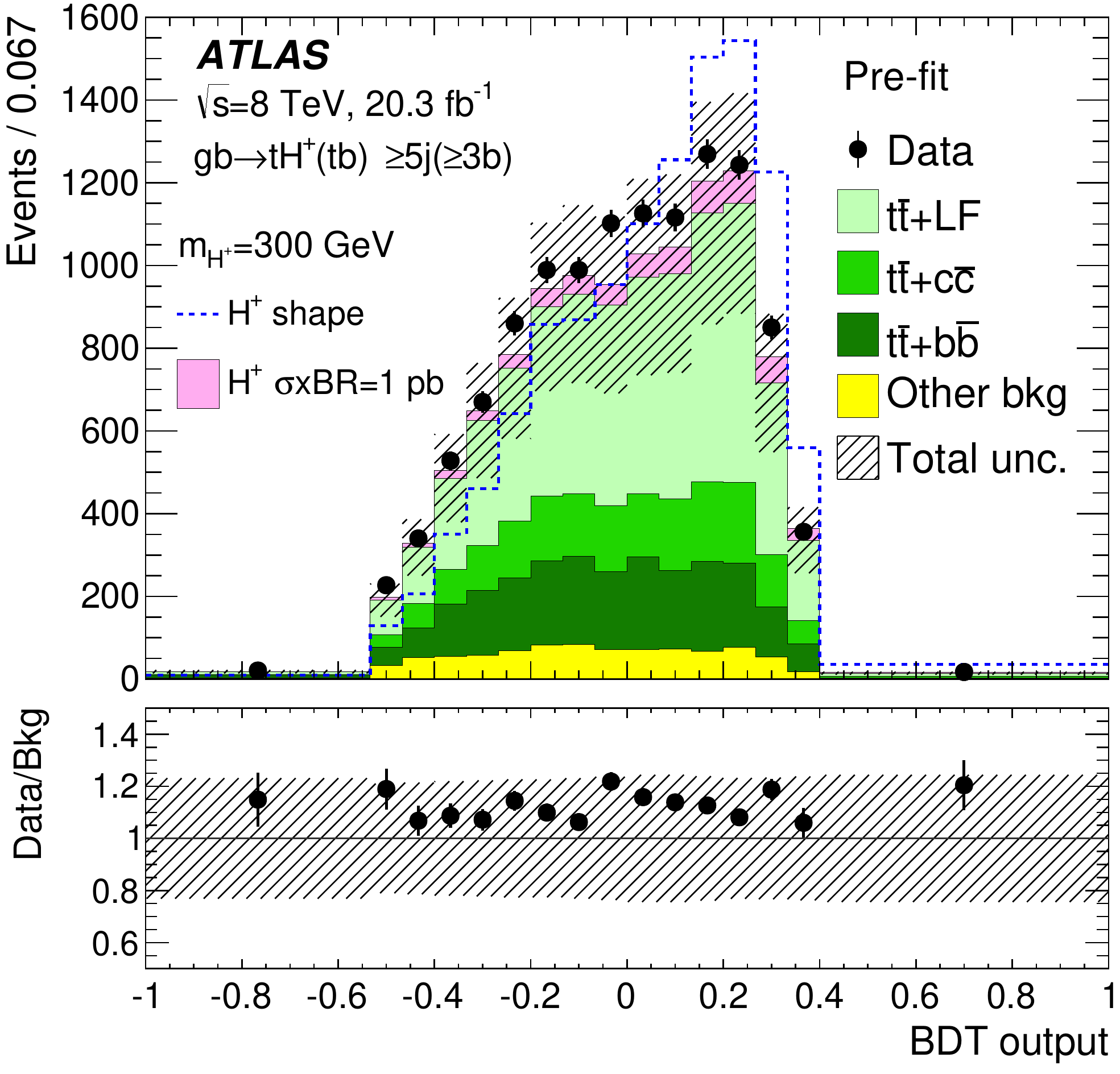} }
\subfloat[]{ \includegraphics[width=0.49\textwidth]{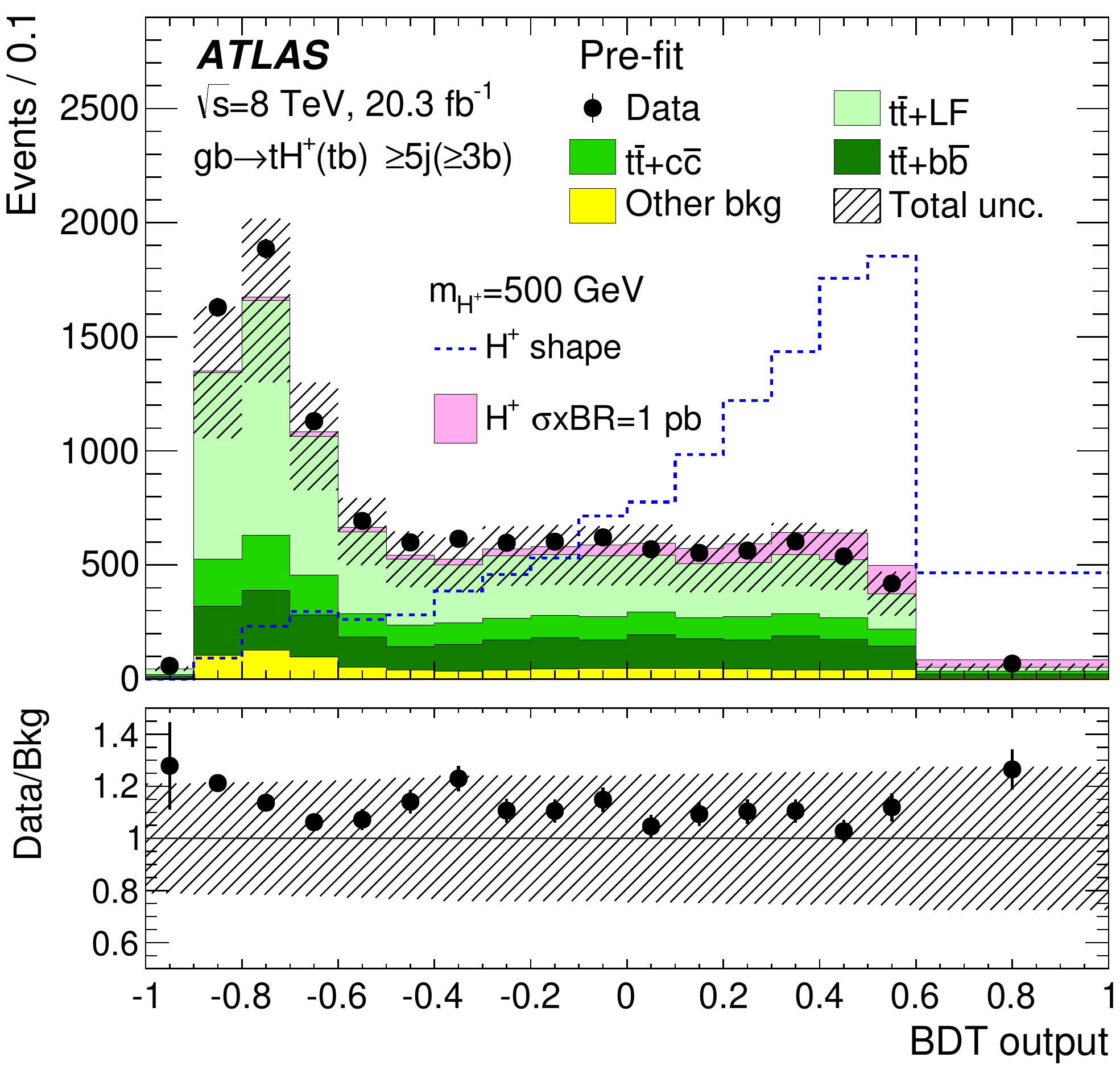} }
\caption{Pre-fit distributions of the BDT output in the signal-rich region trained for two
signal mass hypotheses: (a) 300~\GeV\ and (b) 500~\GeV. Each background process is normalised
according to its cross section. A signal, normalised to a production cross section times
branching fraction for $H^+ \to tb$ ($\sigma \times \mbox{BR}$) of 1~pb, is shown in pink,
stacked on top of the background. The signal shape is shown superimposed as dashed line normalised to the data. The hatched bands show the
pre-fit uncertainties, which are dominated by systematic uncertainties (discussed in
section~\ref{sec:ResolvedSystematics}). The lower panels display the ratio of the data to
the total predicted background. \label{fig:tH_bdtprefit}}
\end{figure}

The pre-fit distributions of $H_\mathrm{T}^\mathrm{had}$ in the four control regions are displayed in figure~\ref{fig:tH_hthadprefit}. Good agreement between data and the SM expectation is found, given the large uncertainties. The pre-fit BDT output distributions for two mass hypotheses are shown in figure~\ref{fig:tH_bdtprefit}. In the SR, the data exceed the expected background, but they are consistent given the large uncertainties. The discrimination between signal and background significantly improves for larger signal masses.

\FloatBarrier

\subsection{Systematic uncertainties}
\label{sec:ResolvedSystematics}

Several sources of systematic uncertainty, affecting the normalisation of signal
and background processes or the shape of their distributions, are considered. The individual
sources of systematic uncertainty are assumed to be uncorrelated, but correlations
of a given systematic effect are maintained across categories and processes, when applicable.
All variations, except those from uncertainties on the theoretical cross section, are symmetrised
with respect to the nominal value. The uncertainties arising from the reconstructed objects and
the background modelling, in particular the $t\bar{t}$ background modelling, receive the same
treatment as in ref.~\cite{tth}.

The following uncertainties on the reconstructed objects are considered. The systematic
uncertainties associated with the electron or muon selection arise from the trigger,
reconstruction and identification efficiency, isolation criteria, as well as from the
momentum scale and resolution~\cite{ElectronPerformance,muons}. In total, the systematic
uncertainties associated with electrons (muons) include five (six) components. The
systematic uncertainties associated with the jet selection arise from the jet energy
scale (JES), the JVF requirement, the jet energy resolution and the jet reconstruction
efficiency. Among these, the JES uncertainty has the largest impact on the search. It is derived
by combining information from test-beam data, LHC collision data and simulation~\cite{JES14060076}.
The JES uncertainty is split into 22 uncorrelated sources, which can have different jet $\pt$- and
$\eta$-dependencies. Six (four) independent sources of systematic uncertainty affecting the
$b$($c$)-tagging efficiency are considered~\cite{btag-data}.
An additional uncertainty is assigned due to the extrapolation of the measurement of
the $b$-tagging efficiency to the high-$\pt$ region. Twelve uncertainties are
considered for the light-jet mistagging rate, with dependencies on the jet
$\pt$ and $\eta$.

The uncertainty on the inclusive $\ttbar$ production cross section is
+5\%/$-$6\%~\cite{Cacciari:2011hy,Beneke,Barnreuther,Czakon1,Czakon2,Czakon3,Czakon:2011xx}.
It accounts for uncertainties from the choice of PDF, $\alphas$ and the top-quark mass.
The PDF and $\alphas$ uncertainties were calculated using the PDF4LHC prescription~\cite{pdf4lhc}
with the MSTW2008 68\% CL NNLO, CT10 NNLO and NNPDF2.3 NNLO~\cite{ffn} PDF sets, added in quadrature
to the scale uncertainty. Systematic uncertainties due to the choice of parton shower and
hadronisation models are derived by comparing $\ttbar$ events produced with {\sc Powheg-Box} interfaced
to either {\sc Pythia} or {\sc Herwig}. Nine uncertainties associated with the experimental measurement
of the $\pt$ of the top quark and the $\ttbar$ system are considered as separate sources of systematic
uncertainty in the reweighting procedure~\cite{top7tev}. Two additional uncorrelated uncertainties are
assigned specifically to $t\bar{t}$+$c\bar{c}$ events, consisting of the full difference between applying
and not applying the $\pt$ reweighting procedure for the top quark and the $\ttbar$ system, respectively.
A conservative systematic uncertainty of 50\% is applied to $t\bar{t}$+$b\bar{b}$ events to account for differences
between the cross sections obtained with {\sc Powheg+Pythia} and the NLO prediction based
on {\sc Sherpa} with {\sc OpenLoops}~\cite{sherpa1,sherpa2}. In the absence of an NLO prediction
for $t\bar{t}$+$c\bar{c}$, the same uncertainty of 50\% is applied to this component
of the $\ttbar$ background. Four additional systematic uncertainties are considered for the
$t\bar{t}$+$c\bar{c}$ background, derived from the simultaneous variation of factorisation and
renormalisation scales, threshold of the parton-jet matching scheme~\cite{MLM}, and $c$-quark mass variations in the simulation
of $\ttbar$ events with {\sc Madgraph+Pythia}, as well as the difference between simulations
of the $t\bar{t}$+$c\bar{c}$ process with {\sc Madgraph+Pythia} and {\sc Powheg+Pythia}. For the
$t\bar{t}$+$b\bar{b}$ background, eight additional systematic uncertainties are considered: three
arise from scale uncertainties, one from the shower recoil model, two from the choice of PDF in
the NLO prediction from {\sc Sherpa} with {\sc OpenLoops} and two from the uncertainties on
multi-parton interaction and final-state radiation, which are not present in {\sc Sherpa}
with {\sc OpenLoops}.

An uncertainty of +5\%/$-$4\% is assumed for the cross section of single-top-quark production~\cite{stopxs1,stopxs2}, corresponding to the weighted average of the theoretical uncertainties on the $s$-, $t$- and $Wt$-channel production modes. One additional systematic uncertainty is considered to account for different ways of handling the interference between $\ttbar$ and $Wt$ events~\cite{Frixione2008}.
For $t\bar{t}V$, an uncertainty of 30\% on the cross section is assumed~\cite{ttV1,ttV2} and an additional uncertainty arises from variations in the amount of radiation.
The uncertainty on the $t\bar{t}H$ cross section is 
+8.9\%/$-$12\%~\cite{xsec2}.
The uncertainties on the $V$+jets and diboson backgrounds are 48\% and 25\%, respectively~\cite{vjetsxs,dibosonxs}. For events with 5 ($\geq$6 jets), one (two) additional uncertainties of 24\% are added in quadrature to account for the extrapolation to higher jet multiplicities. In addition, the full difference between applying and not applying the $\pt$ reweighting for the vector boson is taken as a systematic uncertainty. Uncertainties on the estimate of the multi-jet background come from the limited number of events in the data, especially at high jet and $b$-tagged jet multiplicities, from the uncertainties on the measured lepton misidentification rates (assumed to be 50\%, but uncorrelated between events with an electron or muon), as well as from the subtraction of simulated events with a prompt lepton when estimating the misidentification rates.

Three sources of systematic uncertainty are considered when modelling $H^+ \to tb$ events.
Uncertainties arising from the choice of PDF are estimated using samples generated
with {\sc MC@NLO}~v4.6~\cite{FRI-0201} interfaced to {\sc Herwig++}~v2.5.2~\cite{herwigpp}, by taking
the envelope of the MSTW2008 68\% CL NLO, CT10 NLO and NNPDF2.3 NLO PDF sets,
and by normalising to the nominal cross section~\cite{pdf4lhc}. The uncertainties observed
across the charged Higgs boson mass range are of the order of 5--10\% and increase slightly
with the $H^+$ mass. This systematic uncertainty affects both shape and normalisation.
Uncertainties from the choice of the event generator are estimated from a comparison
of the signal acceptances between events produced using either {\sc Powheg} or
{\sc Madgraph5\_aMC@NLO}~v2.1.1~\cite{Alwall:2014hca}, both interfaced to {\sc Pythia}~v8.1,
with a charged Higgs boson mass of 400~\GeV. In the SR, this uncertainty is found
to be about 1\%, while it increases to as much as 20\% in the CR. It is applied
to all signal mass points as a normalisation-only systematic uncertainty. Uncertainties originating from initial- and final-state parton radiation, which can modify the jet production rate, are evaluated by varying factorisation/renormalisation scale parameters in the production of signal samples. This systematic uncertainty is found to be below 2\% in all five event categories.

\subsection{Results}
\label{sec:ResolvedResults}

A binned maximum likelihood fit to the data is performed simultaneously in the five event categories, and each mass hypothesis is tested separately. The inputs to the simultaneous fit are the distributions of $H_\mathrm{T}^\mathrm{had}$ in the four CR, and the BDT output histograms in the SR. The procedures for quantifying how well the data agree with the background-only hypothesis and for determining exclusion limits are based on the profile likelihood ratio test~\cite{combi2011}. The parameter of interest is the production cross section $\sigma(gb\rightarrow tH^{+})$ multiplied by the branching fraction $\mathrm{BR}(H^{+}\to tb)$, also referred to as the signal strength. All systematic uncertainties, either from theoretical or experimental sources, are implemented as nuisance parameters with log-normal constraint terms. There are about 100 nuisance parameters considered in the fit, the number varying slightly across the range of mass hypotheses. The largest uncertainties for any tested mass point are those arising from the modelling of the $t\bar{t}$ processes. For $m_{H^{+}}<350$~\GeV, the uncertainty on the $t\bar{t}$+$b\bar{b}$ cross section has the largest impact on the result. For higher mass hypotheses, the uncertainties on the shape of the distributions for $t\bar{t}$+$b\bar{b}$ from the reweighting to the NLO prediction are dominant. The fractional contributions of various sources of uncertainty to the total uncertainty on the parameter of interest are presented in table~\ref{tab:tH_sysbreak}, for two hypothesised $H^+$ masses. The uncertainties decrease for higher mass hypotheses because of the larger signal acceptance and the improved separation between signal and background. The pulls of the nuisance parameters after profiling to the data are almost all within $\pm1\sigma$ and never exceed $\pm1.5\sigma$ for all tested mass hypotheses. The pulls that are larger than $\pm1\sigma$ in at least one of the tested mass hypotheses are those associated with uncertainties on the $t\bar{t}$+HF cross sections, on the parton shower modelling of the $t\bar{t}$+$c\bar{c}$ process, and on the $t\bar{t}$+$b\bar{b}$ NLO modelling, derived from variations of the functional form of the renormalisation scale .

\begin{table}[h!]
\begin{center}
\begin{tabular}{c
S[table-format=2.1, table-number-alignment=right]
S[table-format=2.1, table-number-alignment=right]}
\toprule
\multirow{2}{*}{Source of uncertainty} & \multicolumn{2}{c}{Fractional uncertainty [\%]} \\
 & \multicolumn{1}{c}{$m_{H^+}$ = 300 \GeV} & \multicolumn{1}{c}{$m_{H^+}$ = 500 \GeV} \\
\hline
$t\bar{t}$ modelling & 31  & 33 \\
Jets                 & 21  & 9.5 \\
Flavour tagging      & 19  & 24 \\
Other background modelling  & 9.6 & 12 \\
Signal modelling     & 8.0 & 3.5 \\
Lepton     & 1.2 & 0   \\
Luminosity & 1.1 & 0.4 \\
\hline
Statistics & 8.9 & 18  \\
\bottomrule
\end{tabular}
\caption{Percentage of the total uncertainty on the signal strength that is induced from various systematic uncertainties. The values are obtained after fits to the background-plus-signal hypothesis. The largest contribution to the total uncertainty comes from the $t\bar{t}$ modelling. \label{tab:tH_sysbreak}}
\end{center}
\end{table}

\begin{figure}[h!]
  \begin{center}
  \subfloat[]{ \includegraphics[width=0.39\textwidth]{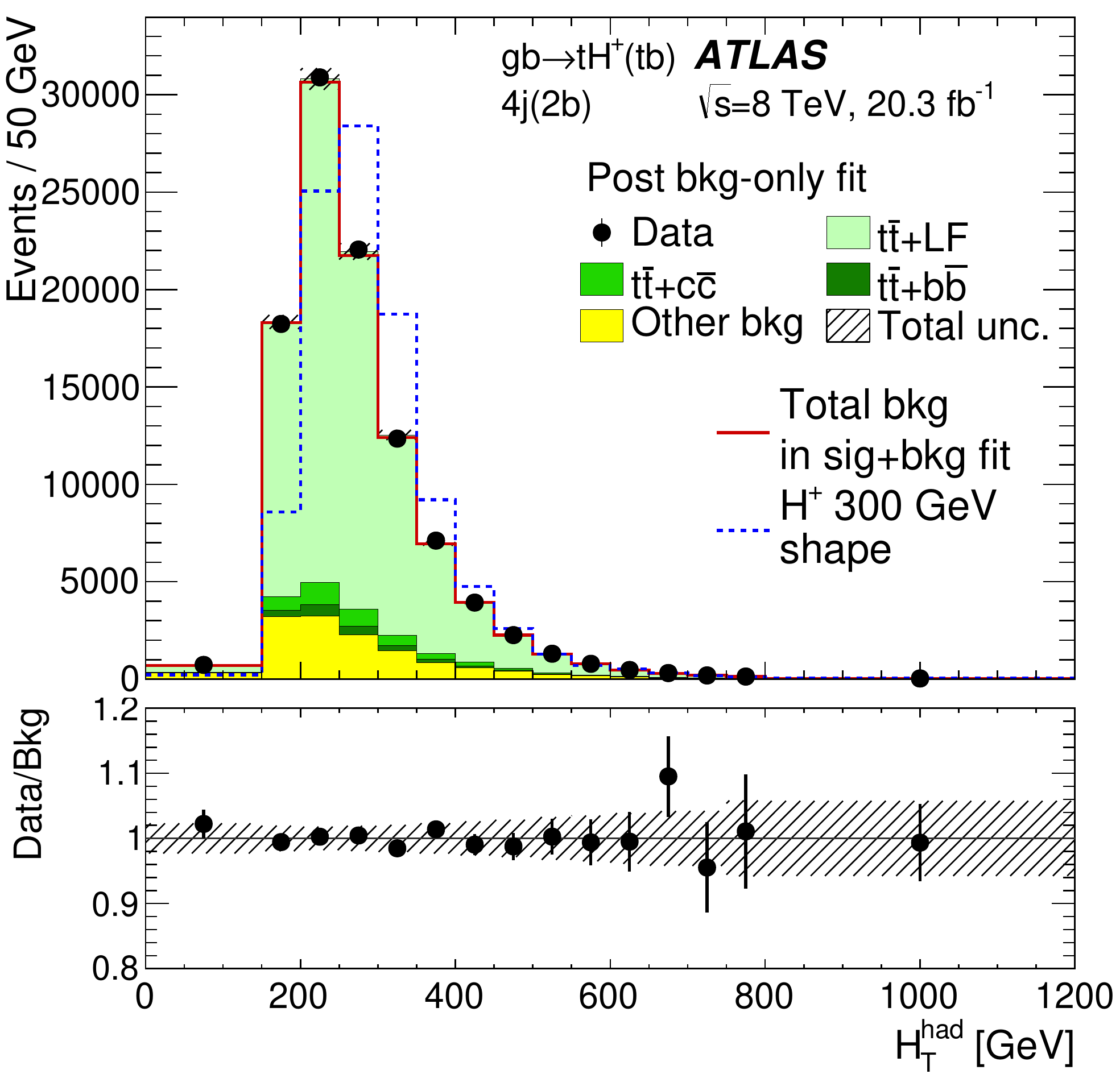} }
  \subfloat[]{ \includegraphics[width=0.39\textwidth]{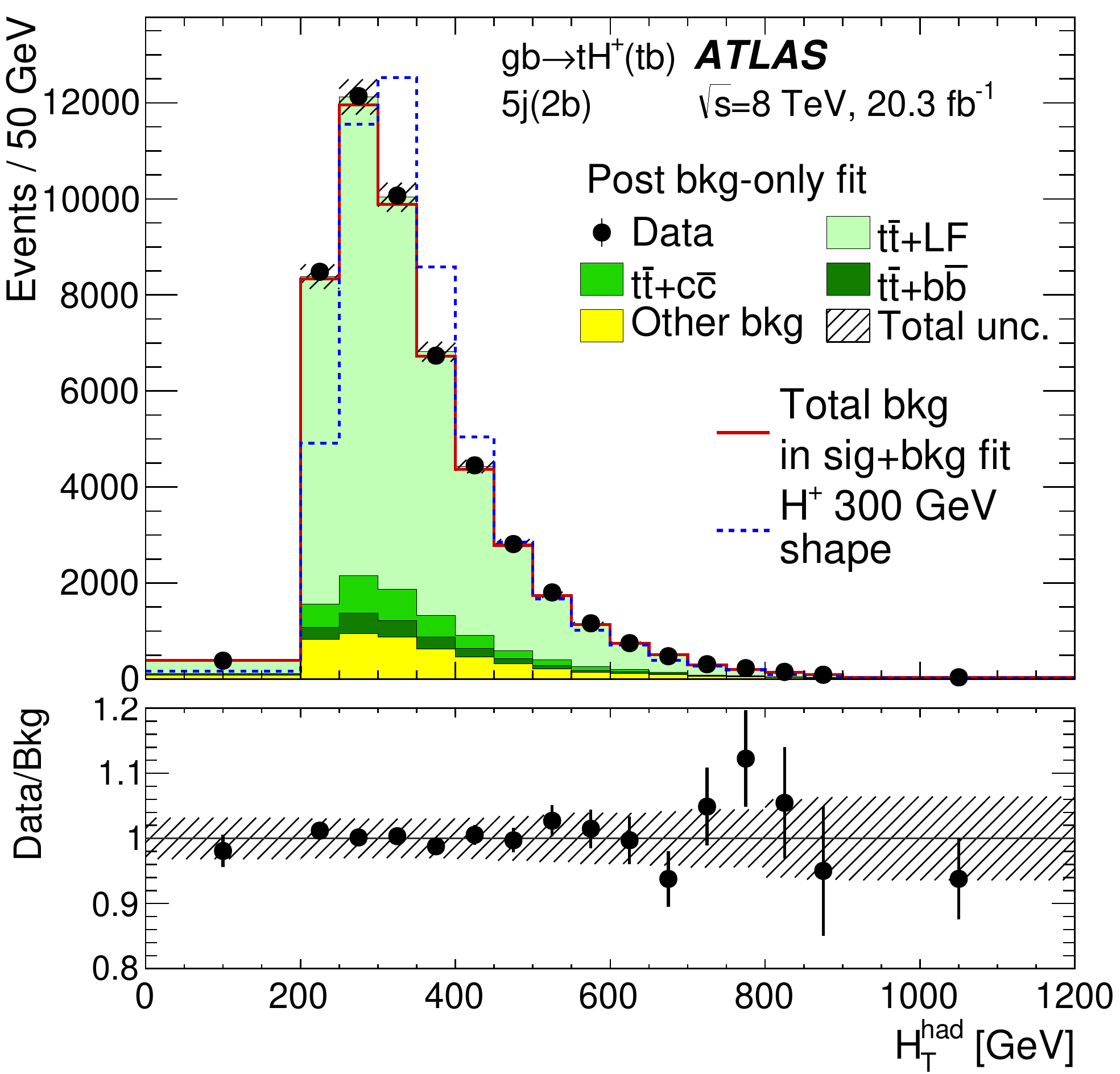} }\\
  \subfloat[]{ \includegraphics[width=0.39\textwidth]{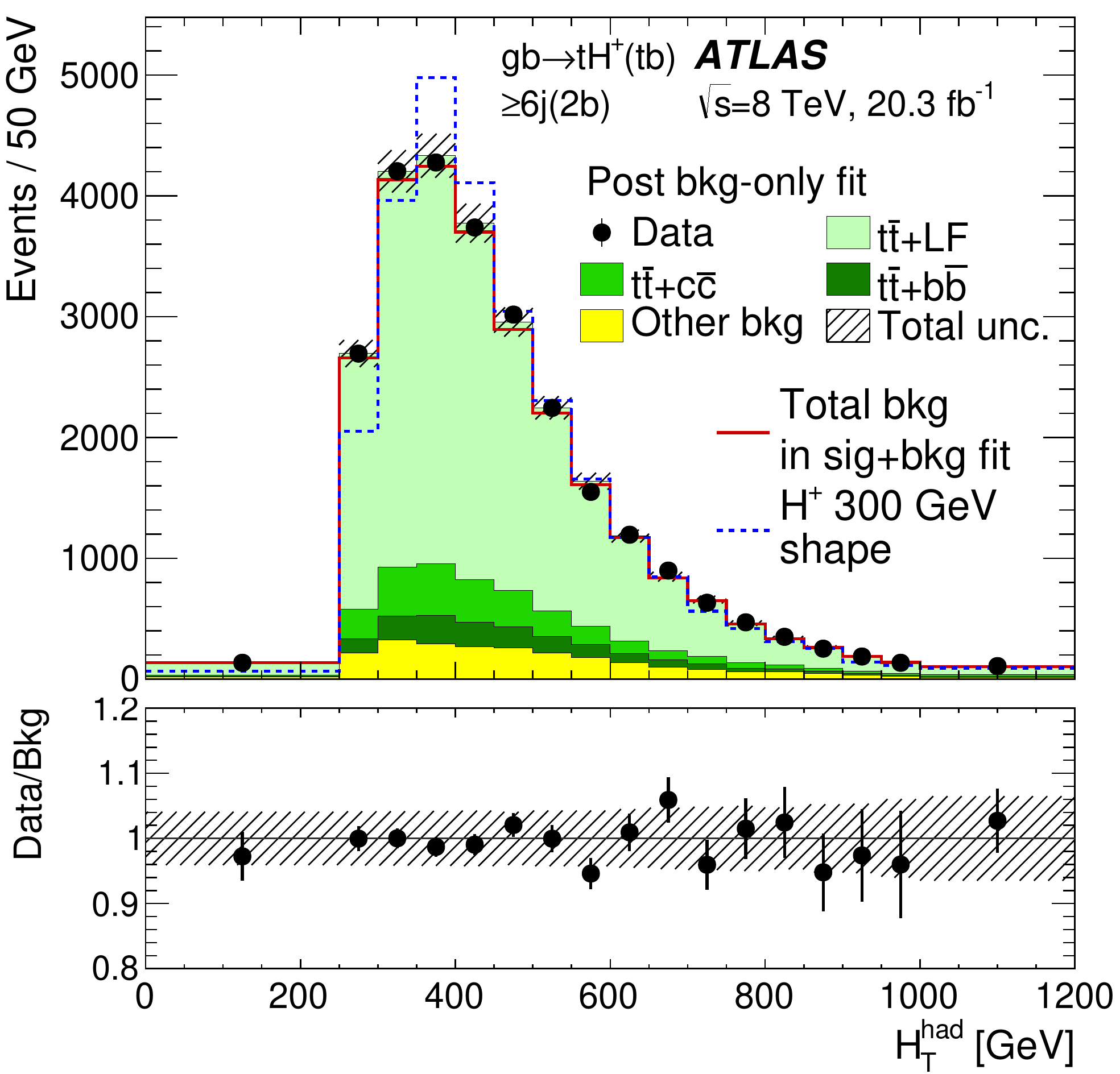} }
  \subfloat[]{ \includegraphics[width=0.39\textwidth]{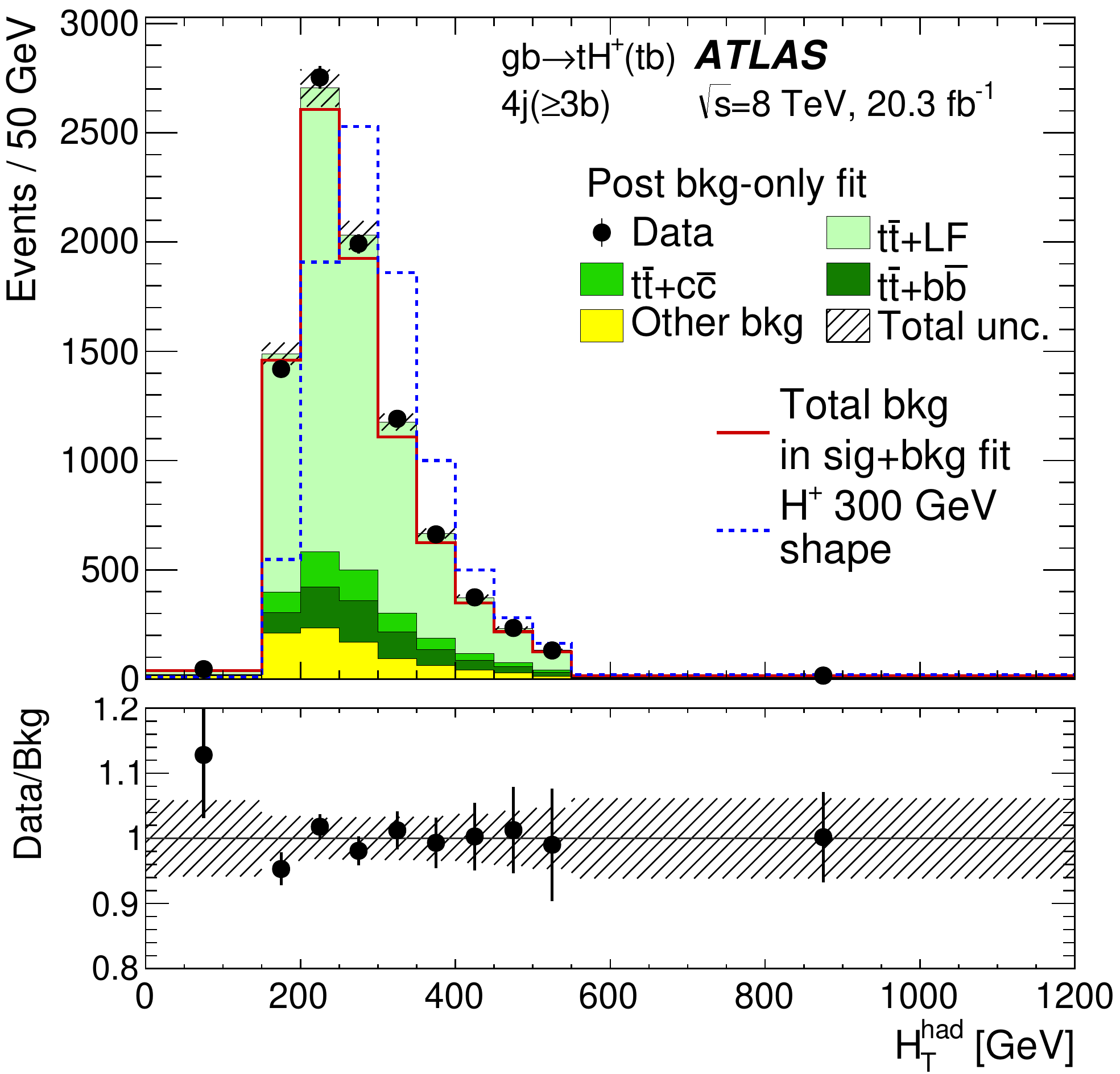} }
  \end{center}
  \caption{Distributions of $H_{\mathrm{T}}^{\mathrm{had}}$ after the fit to the data under the background-only hypothesis in the four control regions: (a) 4j(2b), (b) 5j(2b), (c) $\geq$6j(2b), (d) 4j($\geq$3b). Each background is normalised according to its post-fit cross section. The signal shape is shown as a superimposed dashed blue line normalised to the data. The last bin includes the overflow. The hatched bands show the post-fit uncertainties taking into account the constraints and correlations of the nuisance parameters. The lower panels display the ratio of the data to the total predicted background. In addition, the solid red line shows the total background after an unconditional fit under the background-plus-signal hypothesis with a signal mass of 300~\GeV.\label{fig:tH_hthadpostfit}}
\end{figure}

\begin{figure}[h!]
\centering
\subfloat[]{ \includegraphics[width=0.49\textwidth]{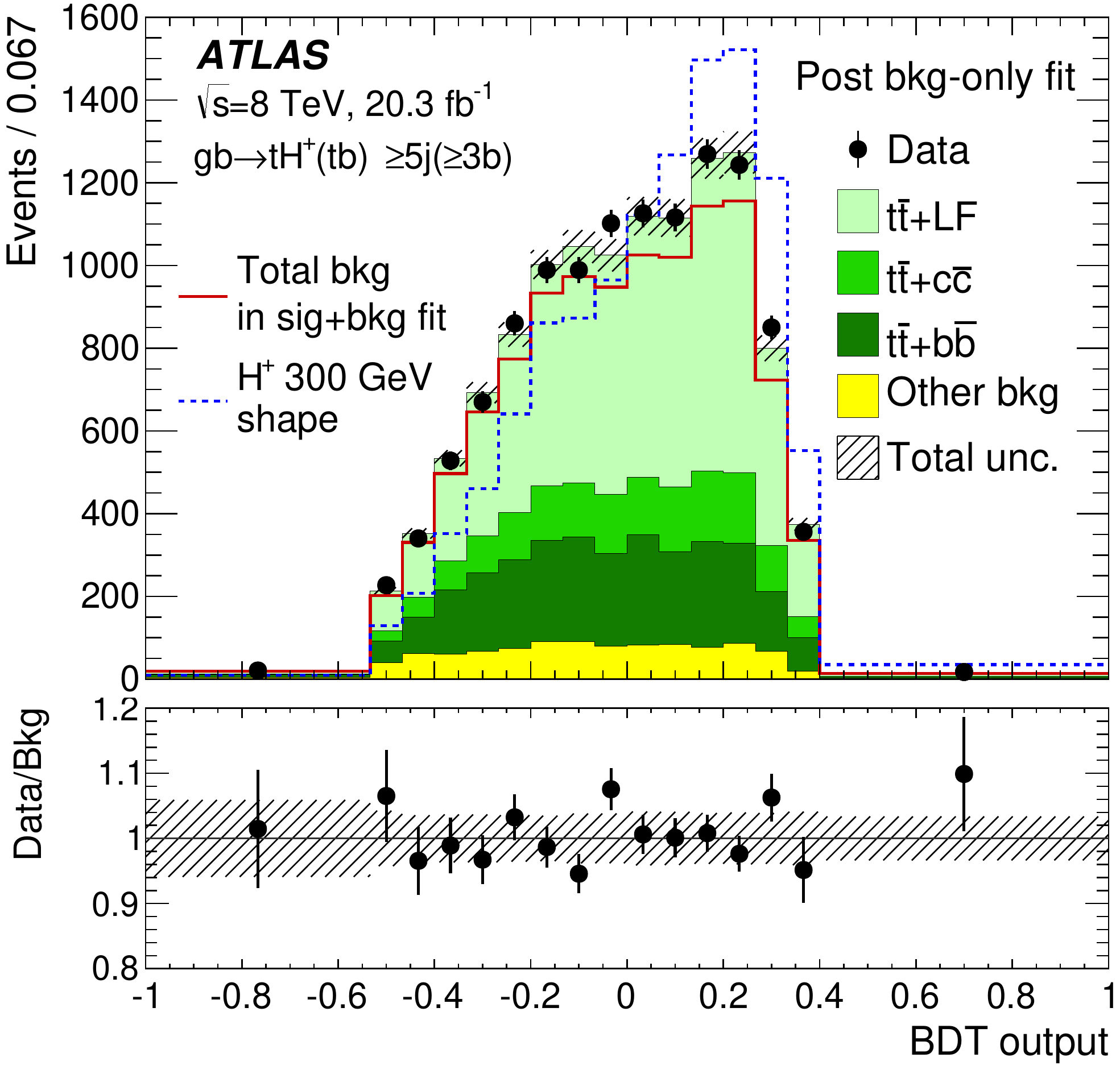} }
\subfloat[]{ \includegraphics[width=0.49\textwidth]{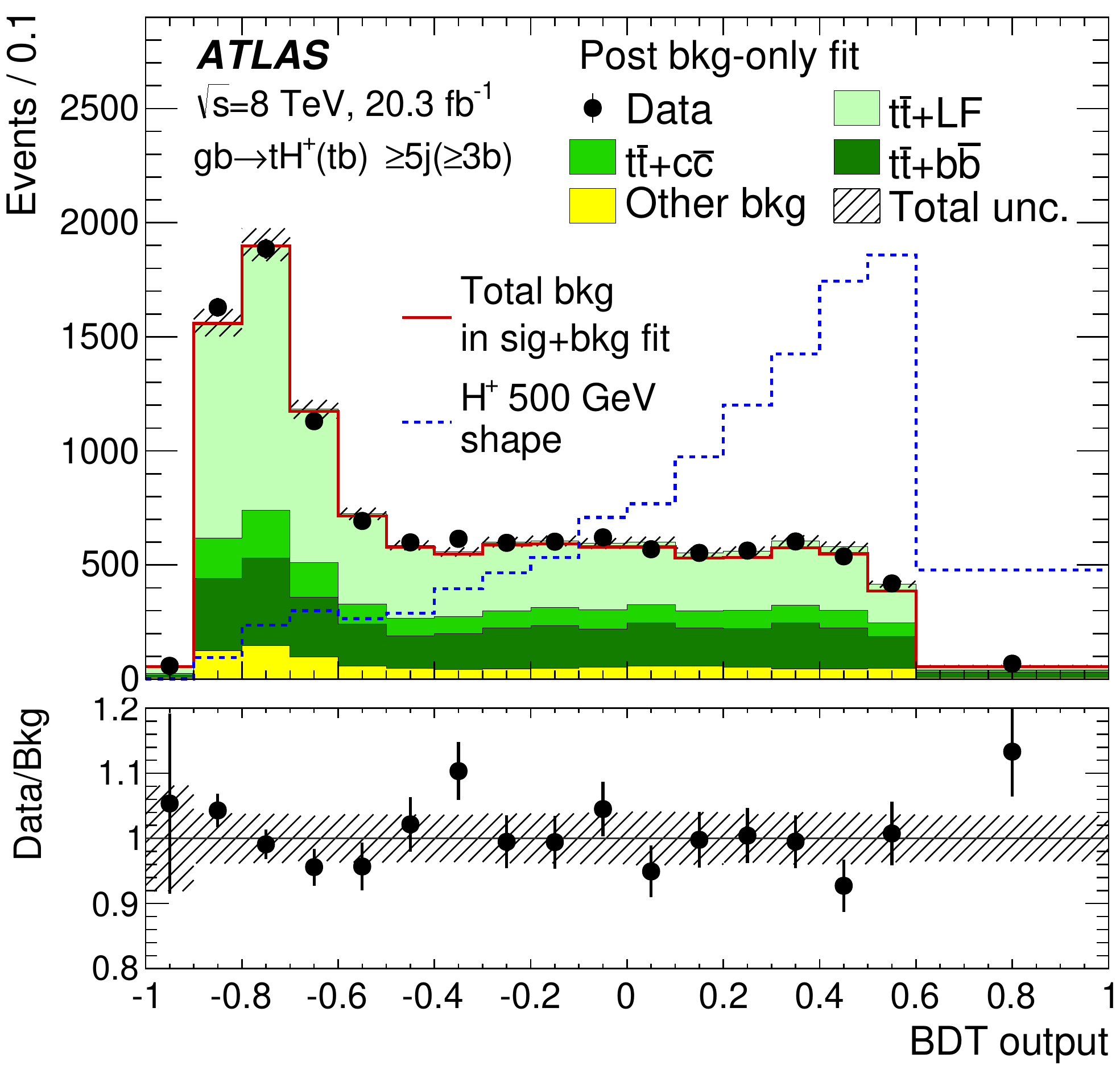} }
\caption{Distributions of the BDT output in the signal-rich region after the fit to the data under the background-only hypothesis. The BDT was trained for two signal mass hypotheses: (a) 300~\GeV\ and (b) 500~\GeV. Each background is normalised according to its post-fit cross section. The signal shape is shown as a superimposed dashed blue line normalised to the data. The hatched bands show the post-fit uncertainties taking into account the constraints and correlations of the nuisance parameters. The lower panels display the ratio of the data to the total predicted background. In addition, the solid red line shows the total background after an unconditional fit under the background-plus-signal hypothesis with a signal mass of (a) 300~\GeV\ and (b) 500~\GeV.\label{fig:tH_bdtpostfit}}
\end{figure}

The post-fit distributions of the $H_\mathrm{T}^\mathrm{had}$ variable in the four CR for the fit under the background-only hypothesis are shown in figure~\ref{fig:tH_hthadpostfit}, whereas the background-only post-fit distributions of the BDT output in the SR are presented in figure~\ref{fig:tH_bdtpostfit}.
The background component of a fit under the background-plus-signal hypothesis is overlayed.
The post-fit event yields for the fit under the background-plus-signal hypothesis for $m_{H^{+}}=300$~\GeV\ are given in table~\ref{tab:tH_postfityields}.
The fit prefers a positive signal strength for all tested mass hypotheses, except at 600~\GeV.
The post-fit event yields for the $t\bar{t}$+HF process are higher in background-only fits than those obtained in fits where the signal hypothesis is included.

\begin{table}[t]
\begin{center}
\resizebox{\textwidth}{!}{
\begin{tabular}{c
S[table-format=6.1, table-number-alignment=right]@{\,}@{$\pm$}@{\,}S[table-format=4.1, table-number-alignment=right]
S[table-format=5.1, table-number-alignment=right]@{\,}@{$\pm$}@{\,}S[table-format=4.1, table-number-alignment=right]
S[table-format=5.1, table-number-alignment=right]@{\,}@{$\pm$}@{\,}S[table-format=4.1, table-number-alignment=right]
S[table-format=4.1, table-number-alignment=right]@{\,}@{$\pm$}@{\,}S[table-format=3.1, table-number-alignment=right]
S[table-format=5.1, table-number-alignment=right]@{\,}@{$\pm$}@{\,}S[table-format=3.1, table-number-alignment=right]}
\toprule
Process & \multicolumn{2}{c}{4j(2b)} & \multicolumn{2}{c}{5j(2b)} & \multicolumn{2}{c}{$\geq$6j(2b)} & \multicolumn{2}{c}{4j($\geq$3b)} & \multicolumn{2}{c}{$\geq$5j($\geq$3b)}\\
\hline
$t\bar{t}$+LF & 83600 & 1900 & 41800 & 1400 & 21000 & 1000 & 6750 & 270 & 6650 & 390  \\
$t\bar{t}$+$c\bar{c}$ & 3200 & 1700 & 2600 & 1400 & 2100 & 1200 & 490 & 230 & 1260 & 570  \\
$t\bar{t}$+$b\bar{b}$ & 1500 & 530  & 1300 & 440  & 1050 & 450  & 600 & 210 & 2040 & 550  \\
$t\bar{t}H$   & 34.6 & 3.8  & 44.6 & 4.9  & 66.7 & 7.8  & 16.2 & 1.9  & 87   & 10  \\
$t\bar{t}V$   & 132  & 39   & 153  & 46   & 186  & 57   & 18.5 & 5.4  & 87   & 26 \\
Single-top    & 5030 & 530  & 1970 & 270  & 860  & 170  & 386  & 55   & 342  & 70  \\
$W$+jets        & 4500 & 1100 & 1660 & 470  & 750  & 270  & 250  & 62   & 220  & 69 \\
$Z$+jets        & 1330 & 560  & 370  & 190  & 137  & 80   & 56   & 23   & 36   & 27 \\
$VV$            & 223  & 63   & 103  & 39   & 47   & 23   & 10.4 & 3.1  & 15.0 & 5.3 \\
Multi-jets    & 2230 & 590  & 690  & 180  & 330  & 100  & 160  & 46   & 208  & 88 \\
\hline
Total bkg & 101800 & 2200 & 50700 & 1600 & 26600 & 1100 & 8730 & 330 & 10950 & 490  \\
\hline
$H^{+}$ & 700 & 310 & 600 & 260 & 430 & 190 & 370 & 160 & 990 & 440  \\
\hline
Data    & \multicolumn{2}{l}{\num{102462}} & \multicolumn{2}{l}{\num{51421}} & \multicolumn{2}{l}{\num{26948}} & \multicolumn{2}{l}{\num{9102}} & \multicolumn{2}{l}{\num{11945}}  \\
\bottomrule
\end{tabular} }
\caption{\label{tab:tH_postfityields} Event yields of SM backgrounds, signal and data in all categories,
after the fit to the data under the background-plus-signal hypothesis with a signal mass of 300~\GeV.
The last column shows the event yields in the SR.
The uncertainties take into account correlations and constraints of the nuisance parameters.}
\end{center}
\end{table}

\begin{figure}[h!]
  \begin{center}
  \includegraphics[width=0.8\textwidth]{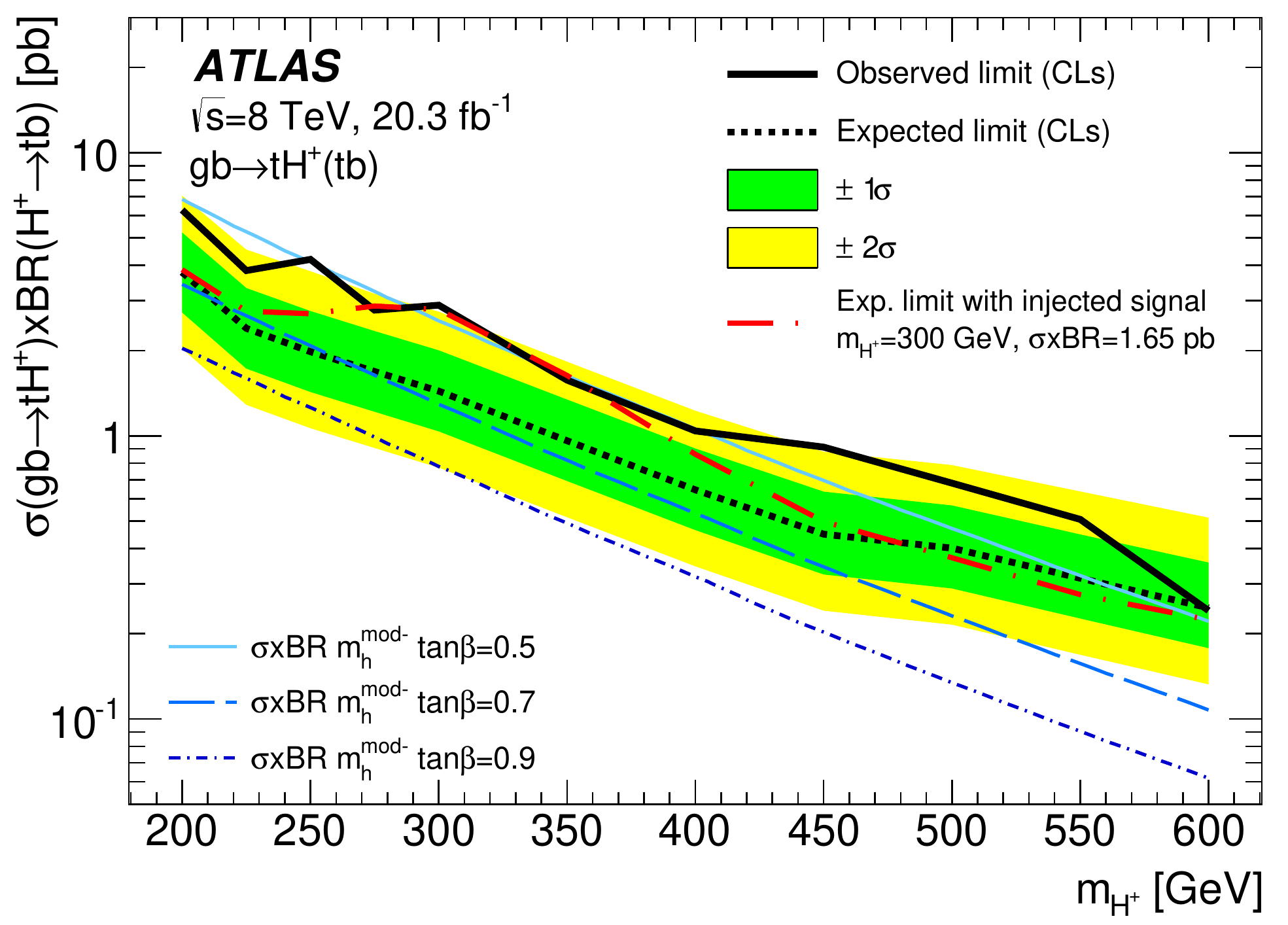}
  \end{center}
  \caption{Expected and observed limits for the production of $H^+ \to tb$ in association with a top quark, as well as bands for 68\% (in green) and 95\% (in yellow) confidence intervals. The red dash-dotted line shows the expected limit obtained in the case where a simulated signal is injected at $m_{H^+} = 300~\mbox{\GeV}$, with a production cross section times branching fraction of 1.65~pb (corresponding to the best-fit signal strength at that mass hypothesis), yielding a deviation from the expectation that extends less to higher and lower mass values than the observed upper limit. Theory predictions are shown for three representative values of $\tan\beta$ in the $m_h^\mathrm{mod-}$ benchmark scenario of the MSSM.\label{fig:tH_limit}}
\end{figure}

The modified frequentist method (CLs)~\cite{CLs} and asymptotic formulae~\cite{Asimov} are used to calculate upper limits on $\sigma(gb\rightarrow tH^{+}) \times \mathrm{BR}(H^{+}\to tb$). The 95\% confidence level (CL) upper limits are presented in figure~\ref{fig:tH_limit}. The mass hypotheses are tested in 25~\GeV\ steps between 200 and 300~\GeV, and in 50~\GeV\ steps up to 600~\GeV.
At 250~\GeV, the local $p_0$-value for the observation to be in agreement with the background-only hypothesis reaches its smallest value of 0.9\% (corresponding to 2.4 standard deviations). At $m_{H^+}$ values of 300 and 450~\GeV, the excess of the data with respect to the background-only hypothesis corresponds to 2.3 standard deviations.

For comparison, the expected upper limit is computed with a signal injected at $m_{H^+} = 300~\mbox{\GeV}$, with a production cross section times branching fraction of 1.65~pb, corresponding to the best-fit value of the signal strength at this mass point.
This results in an excess that is more localised at the injected mass value, i.e. extends less to lower and higher masses than the trend seen in the observed upper limit, as shown in figure~\ref{fig:tH_limit}.
The $H^+$ signal is generated with a zero width. The experimental mass resolution ranges from approximately 30~GeV (for $m_{H^{+}}=200$~GeV) up to 100~GeV (for $m_{H^{+}}=600$~GeV)
and is 50~\GeV\ for the mass hypothesis of 300~\GeV.
A systematic background mismodelling is considerably more likely to give rise to the observed excess than a hypothesised signal at a specific mass.
The cross sections of the $t\bar{t}$+HF backgrounds and the shape
of the $t\bar{t}$+$b\bar{b}$ component have large uncertainties which are correlated with
the signal normalisation. Together with the pre-fit excess of data compared to the SM prediction
(table~\ref{tab:tH_prefityields}), this can result in a post-fit excess over a wide $H^+$ mass range.
The fits were repeated using two alternative, less sensitive, discriminants in the SR: (a) a BDT trained
against the sum of all backgrounds or (b) the variable $H_\mathrm{T}^{\mathrm{had}}$. Similar excesses were observed
with these two alternative methods. The tested mass points are correlated with each other,
since no mass-dependent event selections are applied in the analysis and the dataset is the
same regardless of the hypothesised $\Hplus$ mass. 

The limits in figure~\ref{fig:tH_limit} are presented together with the signal prediction in the $m_h^\mathrm{mod-}$ benchmark scenario of the MSSM~\cite{MSSM}.
Model points with $0.5 \lesssim \tan\beta \lesssim 0.6$ and $\tan\beta \approx 0.5$ are excluded in the $H^+$ mass ranges of 200--300~\GeV\ and 350--400~\GeV, respectively,\footnote{No reliable theoretical predictions exist for $\tan\beta<0.5$.} while the expected limits in the mass range of 200--400~\GeV\ reach $\tan\beta = 0.7$. The $m_h^\mathrm{mod-}$ scenario is chosen as a reference model, but similar exclusions are obtained in other relevant scenarios of the MSSM~\cite{MSSM}, i.e.\ $m_h^\mathrm{mod+}$, $m_h^\mathrm{max-up}$, \emph{tau-phobic}, \emph{light stau} and \emph{light stop}. It has been verified that the width predicted by these models does not have a notable impact on the exclusions.

\clearpage

\section{Search for a charged Higgs boson produced in the $s$-channel}
\label{sec:schannel}

In this section, two searches for $qq' \to \Wprime \to tb$ recently published
by ATLAS~\cite{Aad:2014xea,Aad:2014xra} are reinterpreted as searches for the
$s$-channel production\footnote{While the process generated is $q\qprime \to H^+$,
the most commonly occuring reaction is $cs \to H^+$. }
of charged Higgs bosons, i.e.\ $q\qprime \to H^+ \to tb$,
based on final states with one charged lepton (electron or muon) and jets, or hadronic jets only.

\subsection{Lepton+jets final state}
\label{sec:SchanLplusJets}

In the search for $H^+ \to tb \to (\ell\nu b)b$ produced in the $s$-channel, where the
charged lepton $\ell$ is an electron or muon (from a prompt $W$-boson decay or a leptonic
$\tau$ decay), only events collected using a single-electron or single-muon trigger are
considered, with the same combination of thresholds as in section~\ref{sec:SelectionCutbased}.
Exactly one charged lepton is required, which must match, with $\Delta R < 0.15$, a lepton
reconstructed by the trigger. The electron or muon is then required to have $E_\mathrm{T}$ or $p_\mathrm{T}$ greater than 30~\GeV. The selected events must then have two or
three jets, with exactly two of them $b$-tagged. In addition, the $\met$
must exceed 35~\GeV, and the sum $\met+m_{\mathrm{T}}$, where
$m_{\mathrm{T}}$ is the transverse mass\footnote{The transverse mass is defined as $m_{\mathrm{T}} = \sqrt{2\pt^{\ell}\met(1-\cos\Delta \phi_{\ell,\mathrm{miss}})}$, where $\Delta \phi_{\ell,\mathrm{miss}}$ is the azimuthal separation between the reconstructed lepton and the missing momentum in the transverse plane.}
of the $W$ boson, is required to be greater than 60~\GeV\ in order to reduce the
contribution from the multi-jet background.
Assuming that the missing transverse momentum arises solely from the neutrino in the $W$-boson decay,
its transverse momentum is given by the $x$- and $y$-components of the $\met$ vector, while the
unmeasured $z$-component of the neutrino momentum $p_z^\nu$ is inferred by imposing the $W$-boson
mass constraint on the lepton-neutrino system. This leads to a quadratic equation for $p_z^\nu$.
In the case of two real solutions, the one with the smaller $p_z^\nu$ is chosen.
If the solutions are complex, a real estimate of the $p_z^\nu$ is obtained by a kinematic fit that rescales the neutrino momentum components $p_x^\nu$ and $p_y^\nu$ such that the imaginary term vanishes. The corrected missing transverse momentum of the neutrino is kept as close as possible to the measured $\met$~\cite{stop_tchannel}.

Having determined the four-momentum of the leptonically decaying $W$ boson, the top quark
is then reconstructed. The $b$-tagged jet for which the invariant mass of the $Wb$ system
is closest to $m_{\mathrm{top}}$ is assumed to originate from the top-quark decay, the other $b$-tagged
jet being in turn assigned to the $H^+$ decay. The selected events are then classified into one
signal-rich and one signal-depleted region, separately for events with two or three jets.
The signal-rich region is the subset of the sample with two $b$-tagged jets and an
invariant mass $m_{tb} > 330~\mbox{\GeV}$.
The signal-depleted region is the complementary subset, with two $b$-tagged jets and
$m_{tb} < 330~\mbox{\GeV}$.

The shape and normalisation of the multi-jet background with a misidentified lepton are determined
with the matrix method~\cite{MM}. All other backgrounds are taken from simulation. For $W$+jets events,
the sample composition in the signal-rich and signal-depleted regions with two $b$-tagged jets are similar,
hence an overall renormalisation of the $W$+jets background, based on the event yield measured in
the signal-depleted region, is applied to the events with
two jets. In the events with three jets, the contribution of the $W$+jets background remains below 10\%
and large uncertainties are obtained for the data-driven renormalisation factors, hence the $W$+jets
process with three jets is treated in the same way as the other simulated backgrounds.

\begin{figure}[t]
\begin{center}
\subfloat[]{ \includegraphics[width=0.42\textwidth]{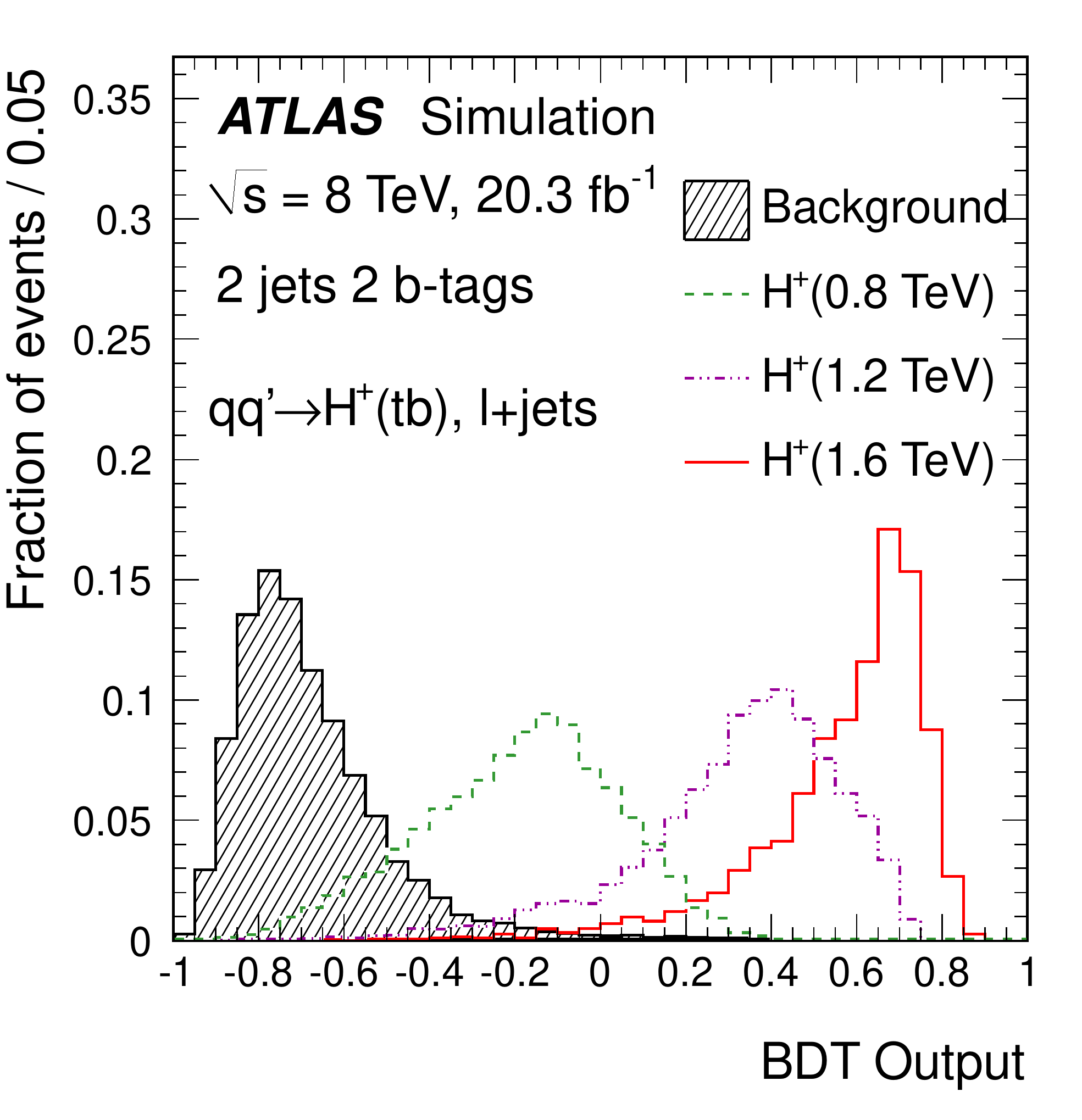} }
\subfloat[]{ \includegraphics[width=0.42\textwidth]{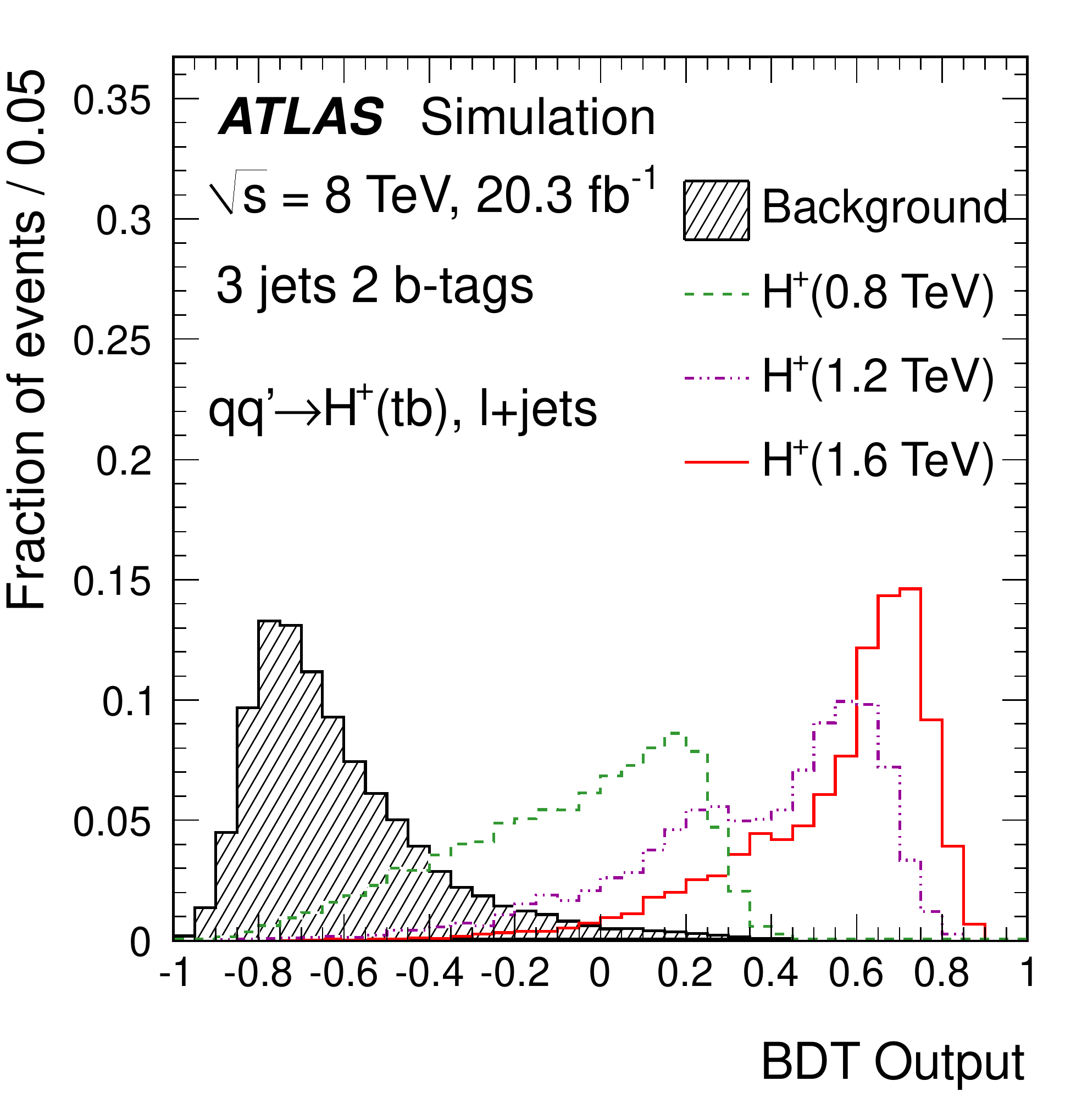} }
\end{center}
\caption{Expected BDT output distribution for the SM backgrounds and for three $H^+$ signal samples
(with masses of 0.8, 1.2 and 1.6~\TeV), obtained in the
signal-rich regions with (a) 2 jets and 2 $b$-tags and (b) 3 jets and 2 $b$-tags.
All distributions are averaged over events with an electron or a muon in the
final state, and they are normalised to unity. \label{fig:BDT-2b} }
\end{figure}

BDT discriminators, using the TMVA toolkit~\cite{tmva}, are again used to
obtain the best separation between the $H^+ \to tb$ signal
events and the associated SM backgrounds.
The same BDT 
as in the search for $\Wprime \to tb$~\cite{Aad:2014xea} are used.
These BDT were trained for a $\Wprime$ mass of 1.75~\TeV.
Alternative trainings with $H^+$ samples were tested but no overall improvement of the expected sensitivity was found.
In events with two
(three) jets, ten (eleven) BDT input variables are used, of which $m_{tb}$ and the $\pt$ of the top-quark
candidate are the most discriminating. Figure~\ref{fig:BDT-2b} shows the expected BDT output
distributions, normalised to unity, for selected $H^+ \to tb$ signal samples and the background processes,
in the signal-rich regions.

\begin{figure}[t]
\begin{center}
\subfloat[]{ \includegraphics[width=0.42\textwidth]{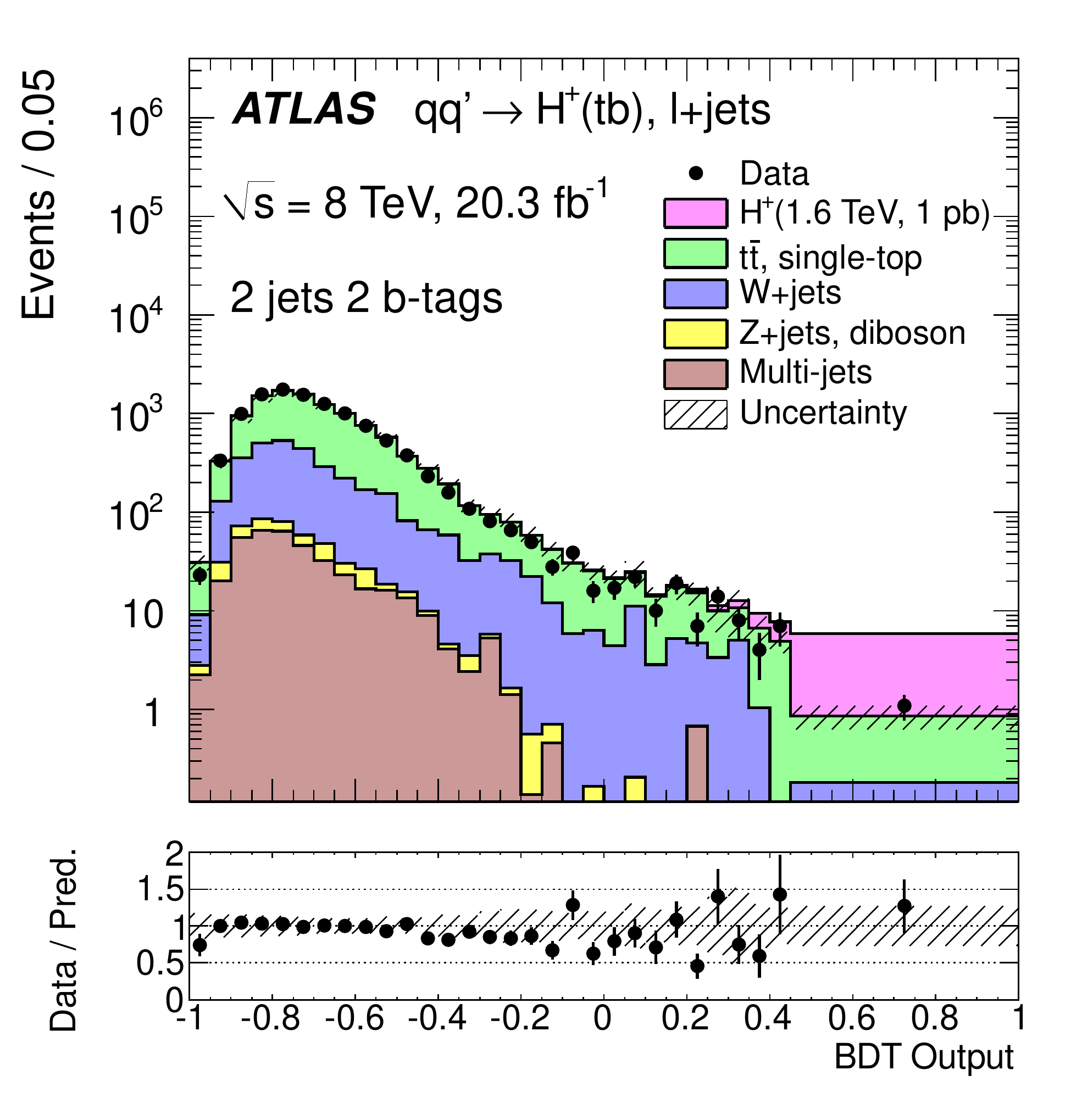} }
\subfloat[]{ \includegraphics[width=0.42\textwidth]{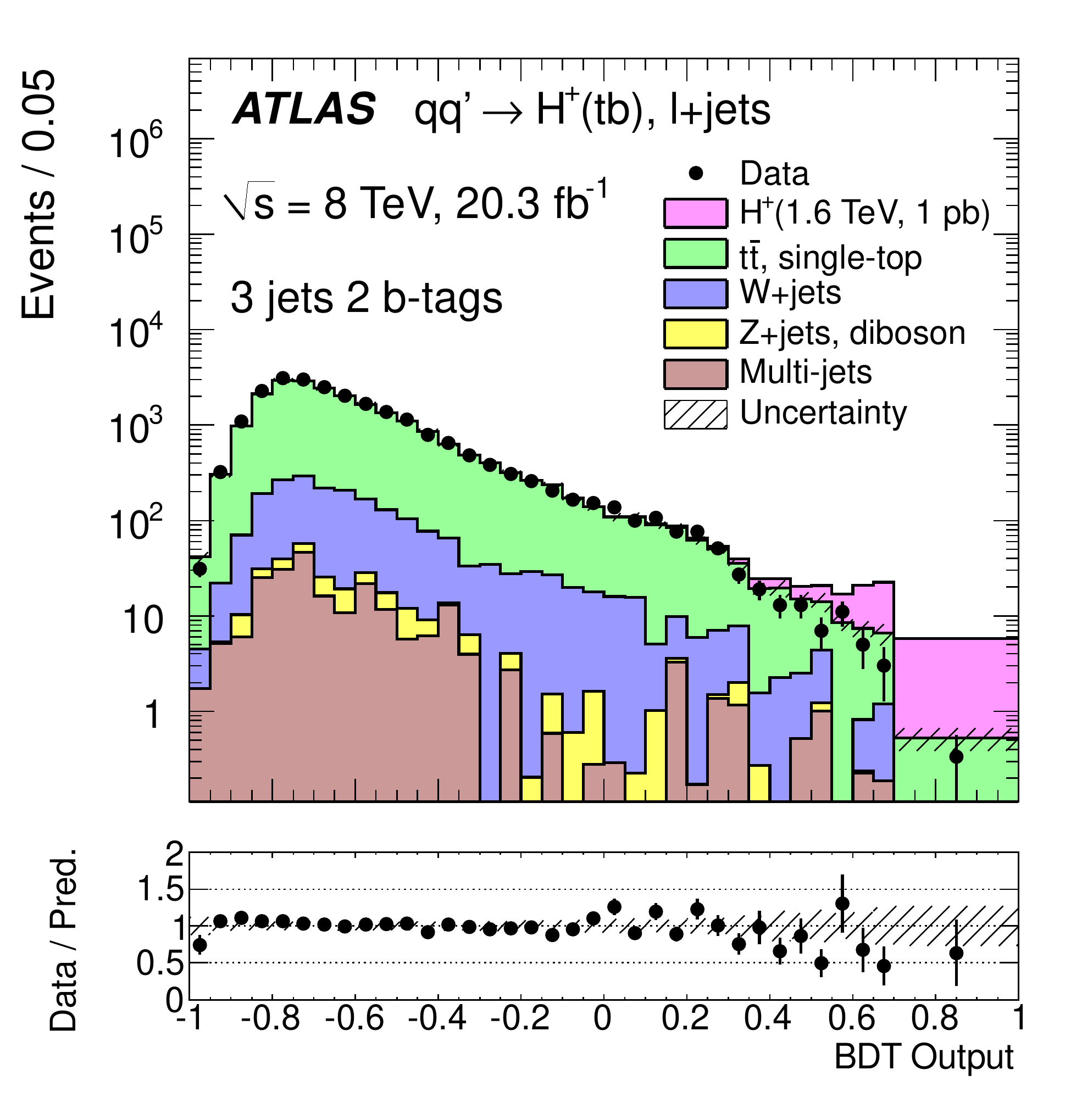} }
\end{center}
\caption{Comparison of the BDT output distributions between the ATLAS data and simulation,
in the signal-rich regions with (a) 2 jets and 2 $b$-tags and (b) 3 jets and 2 $b$-tags,
summing the events
with an electron or a muon in the final state. A potential signal contribution, with a charged
Higgs boson mass of 1.6~\TeV\ and a cross section times branching fraction of 1~pb, is shown
stacked on top of the background distributions.
The uncertainty bands include normalisation uncertainties on all backgrounds
and the uncertainty due to the limited size of the samples of simulated events. \label{fig:BDT-2b-data} }
\end{figure}

No sign of a signal is observed in the selected samples with two or three jets, including two $b$-tags~\cite{Aad:2014xea}, as illustrated in
figure~\ref{fig:BDT-2b-data}. The BDT distributions of events with 2-jet and 3-jet
final states, with separated $e$+jets and $\mu$+jets samples, are used in a combined
statistical analysis to compute exclusion limits on the cross section times branching fraction for $H^+ \to tb$
in the $s$-channel production mode, as discussed in section~\ref{sec:ResultsSchannel}.

\subsection{All-hadronic final state}
\label{sec:SchanAllHad}

In this section, the search for $H^+ \to tb \to (qq'b)b$ produced in the $s$-channel is described. The selection and the statistical analysis are identical to those of the $\Wprime \to tb$ search~\cite{Aad:2014xra}.
Events with isolated charged leptons are vetoed in the event selection.
Candidate events are first collected using the requirement that the scalar sum of $E_{\mathrm{T}}$
for all energy deposits in the calorimeters exceeds
700~\GeV\ at the trigger level. Then, the scalar sum of $\pt$ of all small-radius jets
is required to be greater than 850~\GeV. The selected
events must contain exactly one top-tagged large-radius jet (reconstructed and identified using the
procedure described in section~\ref{sec:toptagging}) with $\pt > 350$~\GeV\ and $|\eta|<2.0$.
A small-radius $b$-tagged jet, with $\pt > 350$~\GeV\ and a separation $\Delta R > 2.0$
from the top-tagged jet, is also required.
The invariant mass of the top-tagged jet and the $b$-tagged jet, $m_{tb}$, must exceed 1.1~\TeV.
The selected events are classified into two categories, one $b$-tag or two $b$-tags, depending
on whether or not an additional
small-radius $b$-tagged jet with $\pt > 25$~\GeV\ is found with a distance $\Delta R < 1.0$
from the top-tagged jet.
The second $b$-tagged jet, if found, is used for classification only and does not enter
the invariant mass calculation, to avoid double-counting of energy.

\begin{figure}[ht]
\begin{center}
\subfloat[]{ \includegraphics[width=0.44\textwidth]{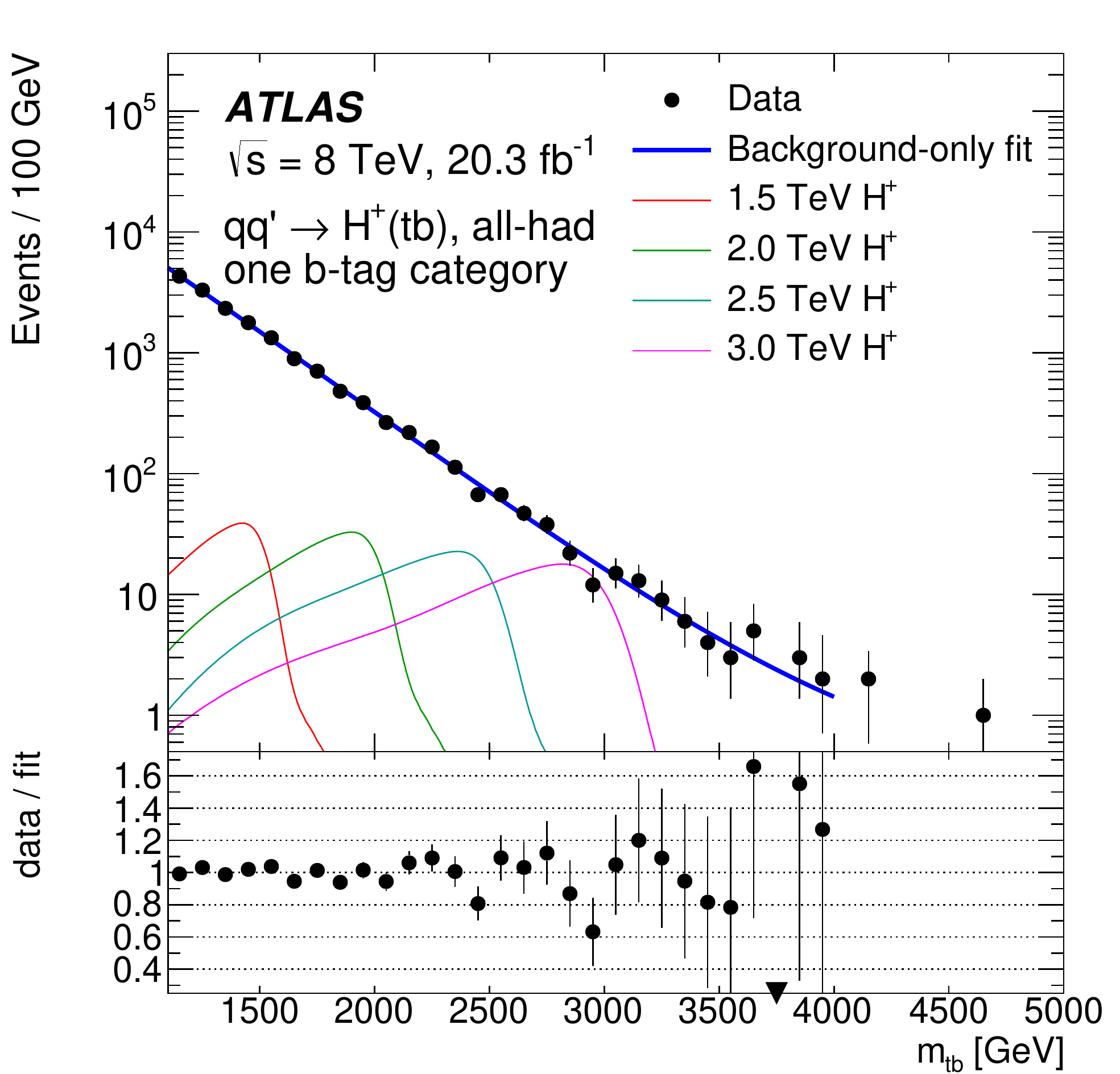} }
\subfloat[]{ \includegraphics[width=0.44\textwidth]{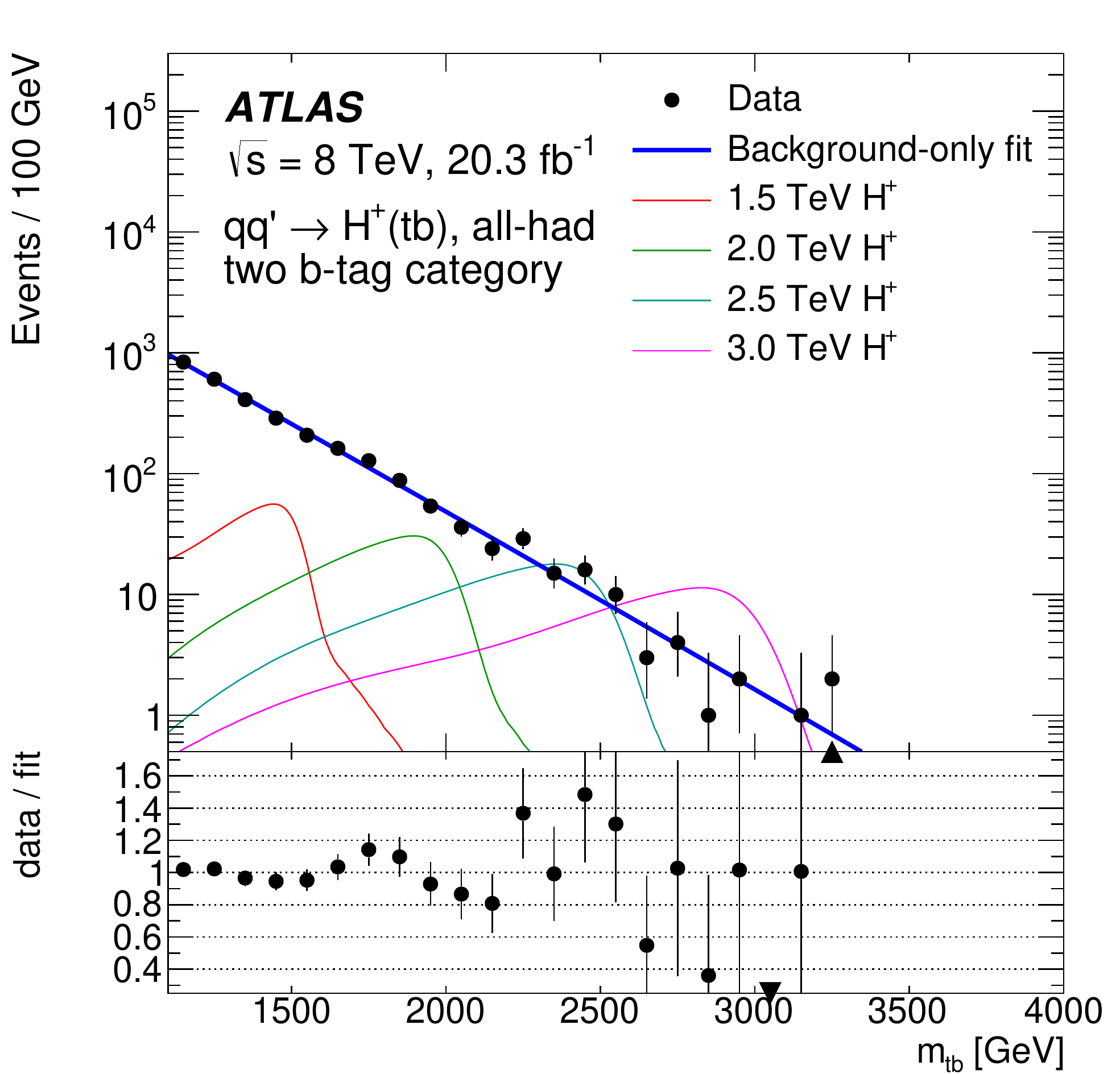} }
\end{center}
\caption{The $m_{tb}$ distribution in data, with a background-only fit, in the
(a) one $b$-tag and (b) two $b$-tag categories.
The lower panels show the ratio of the data to the fit. Potential signal contributions,
with charged Higgs boson masses of 1.5, 2.0, 2.5 and 3.0~\TeV, each corresponding to a cross section
times branching fraction of 0.2~pb, are also shown. \label{fig:tb-allhad-data} }
\end{figure}

The shape of the $m_{tb}$ distribution for the signal is estimated from a fit to
simulated $H^+$ events.
The appropriate functional form is found to be the same as in the search for $\Wprime \rightarrow tb$:
a skew-normal distribution convolved with a Gaussian function,
to capture the asymmetric structure of the $H^+$ signal shape due to radiation,
together with off-shell production~\cite{Aad:2014xra}. The signal shapes are shown
in figure \ref{fig:tb-allhad-data}.

A fit of the SM background plus the $\Hplus$ signal shape to the data is used to estimate the background.
The background shape is described by an exponential function with a polynomial of order $n$ as argument,
$\exp(\sum_{k=1}^n c_k m_{tb}^k)$ with $n=4 \, (2)$ in the one (two) $b$-tag category.
The function was selected to optimally describe the SM background as estimated from fits to signal-free control regions, as well as to minimise the number of spurious signal events found in the background-plus-signal fit to this background-only sample.
Multi-jet events contribute at the level of 99\% (88\%) to the total background in the one (two)
$b$-tag event categories, as estimated from simulation and fits to the data in control
regions~\cite{Aad:2014xra}.

No significant excess of data with respect to the SM predictions is observed in the selected samples with one or two $b$-tags, as shown in
the search for $\Wprime \to tb$~\cite{Aad:2014xra} and illustrated in figure~\ref{fig:tb-allhad-data}.
The $m_{tb}$ distributions in the one and two $b$-tag event categories are used in a
combined statistical analysis to compute exclusion limits, as discussed in section~\ref{sec:ResultsSchannel}.

\subsection{Results and interpretations}
\label{sec:ResultsSchannel}

\begin{figure}[t]
\begin{center}
\subfloat[\label{F:limits-s-chan_ljets}]{ \includegraphics[width=0.45\textwidth]{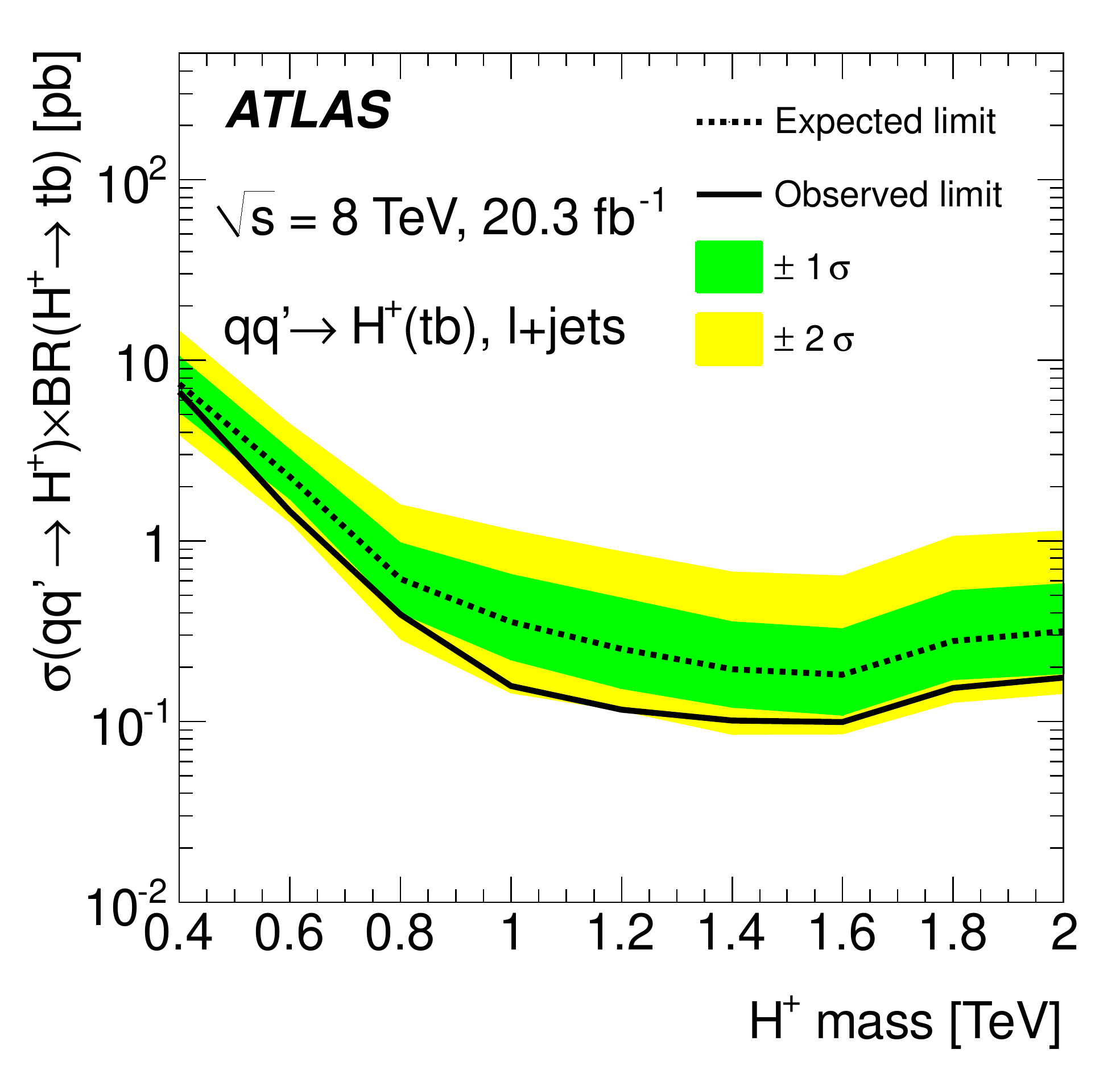} }
\subfloat[\label{F:limits-s-chan_allhad}]{ \includegraphics[width=0.45\textwidth]{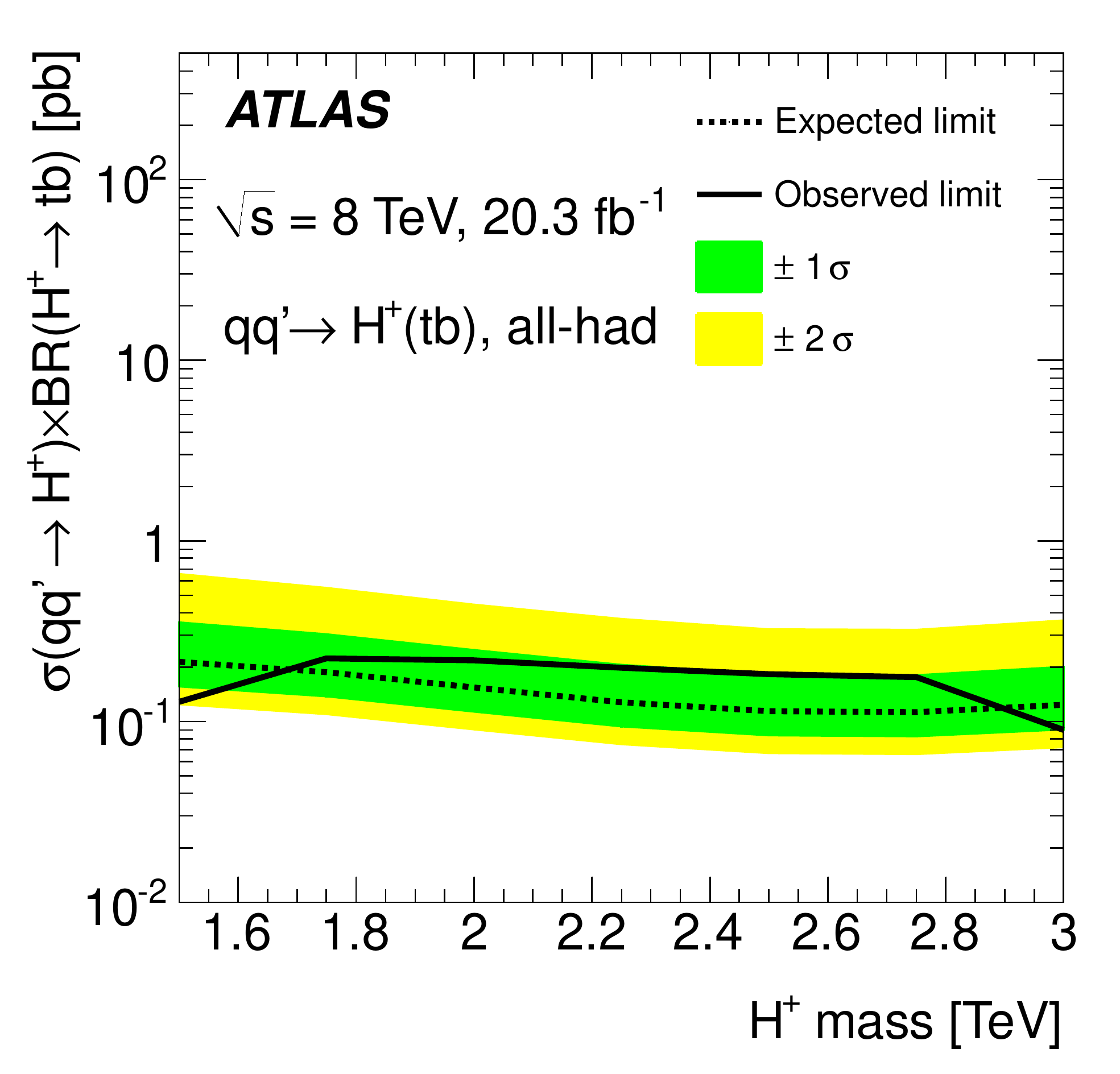} }
\end{center}
\caption{Expected and observed 95\% CL limits on the $s$-channel production cross
section times branching fraction for $H^+ \to tb$ as a function of the charged
Higgs boson mass, in the (a) lepton+jets final state and (b) all-hadronic final state,
including all systematic uncertainties, using a narrow-width approximation.}
\label{fig:limits-s-chan}
\end{figure}

The data are found to be compatible with the background-only predictions~\cite{Aad:2014xea,Aad:2014xra},
and 95\% CL upper limits on the production cross section times branching fraction of $H^+ \to tb$ in the $s$-channel
are derived using a narrow-width approximation. Hypothesis testing is performed using the
CLs~\cite{CLs} procedure,
with the log-likelihood ratio of the background-plus-signal and background-only hypotheses
as the test statistic for both final states. Systematic uncertainties are treated
as nuisance parameters and are implemented in the same manner as in the searches for
$\Wprime \to tb$~\cite{Aad:2014xea,Aad:2014xra}, with the exception of the uncertainty
arising from the choice of the PDF in the signal modelling, since
the colliding partons are mainly $c$- and $s$- quarks in the $H^+$ production.
The PDF systematic uncertainties are estimated by taking the envelope
of the MSTW2008 68\% CL NLO, CT10 NLO and NNPDF3.0 NLO PDF sets in nominal $H^+$ signal events,
reweighted using {\sc LHAPDF6}~\cite{lhapdf}.
The dominant systematic uncertainty in the lepton+jets final state is the $W$+jets
cross section normalisation, while for the all-hadronic final state, the $b$-tagging
and background modelling uncertainties dominate. Figure~\ref{fig:limits-s-chan} shows the
expected and observed 95\% CL upper limits on the production cross section times branching
fraction of $qq' \to H^+ \to tb$ in the $s$-channel. For the lepton+jets (all-hadronic)
final state and the charged Higgs boson mass range of 0.4--2.0~\TeV\ (1.5--3.0~\TeV), these
observed upper limits lie between 0.13 and 6.7~pb (0.09 and 0.22~pb). The corresponding expected
upper limits on the cross section times branching fraction are 0.18--7.4~pb (0.11--0.21~pb).
These limits are valid for a narrow-width approximation, i.e.\ when the decay width divided by
the mass is small ($\Gamma(\Hplus \rightarrow tb)/m_{\Hplus} <1.5\%$) compared with the detector resolution ($\sim$10\%).

No exclusion of a type-II 2HDM in a narrow-width approximation can be made based on the observed limits.
However, these generic upper limits are the first ones from ATLAS for a narrow charged scalar particle produced
through annihilation of light quarks and decaying into a $tb$ pair.
This could enable the probing of charged Higgs bosons (other than type-II 2HDM) that also have sizeable couplings to lighter quarks.

\FloatBarrier
\section{Conclusions}
\label{sec:Conclusion}

\FloatBarrier

This paper presents searches for charged Higgs bosons decaying through $H^+ \to tb$, produced either
in association with a top quark or in the $s$-channel process $qq' \to H^+ \to tb$, using the 20.3~\ifb\ dataset of $pp$
collisions at $\sqrt{s} = 8$~\TeV\ collected by the ATLAS experiment at the LHC during Run 1.

The search for $gb \to tH^+$ is performed in the $H^+$ mass range of 200--600~\GeV.
The analysis uses multivariate analysis techniques in the signal-rich region, and it employs control regions to reduce the large uncertainties on the backgrounds.
An excess of data with respect to the SM predictions is observed for all $H^+$ mass hypotheses, except 600~\GeV.
The injection of simulated $H^+$ events yields a deviation from the expectation that extends less to higher and lower masses than the observed upper limit, indicating that a systematic background mismodelling is more likely to give rise to the observed excess than a signal.
The smallest local $p_0$-values are found at $m_{H^{+}}$ values of 250, 300 and 450~\GeV, corresponding to 2.3--2.4 standard deviations.
The $m_h^{\mathrm{mod-}}$ scenario of the Minimal Supersymmetric Standard Model is excluded at 95\% confidence level for $0.5 \lesssim \tan\beta \lesssim 0.6$ in the $H^+$ mass range of 200--300~\GeV, and for $\tan\beta \approx 0.5$ in the $H^+$ mass range of 350--400~\GeV.

The $s$-channel production of $qq' \to H^+ \to tb$ is investigated through a reinterpretation of searches for $\Wprime \to tb$ in ATLAS.
The lepton+jets final state is used for $H^+$ masses between 0.4 and 2.0~\TeV, and the search employs multivariate techniques in order to reduce the contribution of SM backgrounds.
The all-hadronic final state is used in the $H^+$ mass range of 1.5--3.0~\TeV, and events with a jet tagged as originating from a hadronic top-quark decay are selected in the analysis.
In both searches for $H^+ \to tb$ produced via the $s$-channel process, no significant excess of data is observed with respect to the SM predictions.
The $s$-channel production mode offers a possibility to probe the coupling between light quarks and a charged Higgs boson.
No upper limits on the cross section of charged scalar particles in the $s$-channel production mode have been set previously by the ATLAS experiment.

\section*{Acknowledgements}

We thank CERN for the very successful operation of the LHC, as well as the
support staff from our institutions without whom ATLAS could not be
operated efficiently.

We acknowledge the support of ANPCyT, Argentina; YerPhI, Armenia; ARC, Australia; BMWFW and FWF, Austria; ANAS, Azerbaijan; SSTC, Belarus; CNPq and FAPESP, Brazil; NSERC, NRC and CFI, Canada; CERN; CONICYT, Chile; CAS, MOST and NSFC, China; COLCIENCIAS, Colombia; MSMT CR, MPO CR and VSC CR, Czech Republic; DNRF, DNSRC and Lundbeck Foundation, Denmark; IN2P3-CNRS, CEA-DSM/IRFU, France; GNSF, Georgia; BMBF, HGF, and MPG, Germany; GSRT, Greece; RGC, Hong Kong SAR, China; ISF, I-CORE and Benoziyo Center, Israel; INFN, Italy; MEXT and JSPS, Japan; CNRST, Morocco; FOM and NWO, Netherlands; RCN, Norway; MNiSW and NCN, Poland; FCT, Portugal; MNE/IFA, Romania; MES of Russia and NRC KI, Russian Federation; JINR; MESTD, Serbia; MSSR, Slovakia; ARRS and MIZ\v{S}, Slovenia; DST/NRF, South Africa; MINECO, Spain; SRC and Wallenberg Foundation, Sweden; SERI, SNSF and Cantons of Bern and Geneva, Switzerland; MOST, Taiwan; TAEK, Turkey; STFC, United Kingdom; DOE and NSF, United States of America. In addition, individual groups and members have received support from BCKDF, the Canada Council, CANARIE, CRC, Compute Canada, FQRNT, and the Ontario Innovation Trust, Canada; EPLANET, ERC, FP7, Horizon 2020 and Marie Skłodowska-Curie Actions, European Union; Investissements d'Avenir Labex and Idex, ANR, Region Auvergne and Fondation Partager le Savoir, France; DFG and AvH Foundation, Germany; Herakleitos, Thales and Aristeia programmes co-financed by EU-ESF and the Greek NSRF; BSF, GIF and Minerva, Israel; BRF, Norway; the Royal Society and Leverhulme Trust, United Kingdom.

The crucial computing support from all WLCG partners is acknowledged
gratefully, in particular from CERN and the ATLAS Tier-1 facilities at
TRIUMF (Canada), NDGF (Denmark, Norway, Sweden), CC-IN2P3 (France),
KIT/GridKA (Germany), INFN-CNAF (Italy), NL-T1 (Netherlands), PIC (Spain),
ASGC (Taiwan), RAL (UK) and BNL (USA) and in the Tier-2 facilities
worldwide.

\printbibliography


\clearpage
\begin{flushleft}
{\Large The ATLAS Collaboration}

\bigskip

G.~Aad$^\textrm{\scriptsize 85}$,
B.~Abbott$^\textrm{\scriptsize 113}$,
J.~Abdallah$^\textrm{\scriptsize 151}$,
O.~Abdinov$^\textrm{\scriptsize 11}$,
R.~Aben$^\textrm{\scriptsize 107}$,
M.~Abolins$^\textrm{\scriptsize 90}$,
O.S.~AbouZeid$^\textrm{\scriptsize 158}$,
H.~Abramowicz$^\textrm{\scriptsize 153}$,
H.~Abreu$^\textrm{\scriptsize 152}$,
R.~Abreu$^\textrm{\scriptsize 116}$
% ...
\bigskip
\\
% ...
$^{11}$ Institute of Physics, Azerbaijan Academy of Sciences, Baku, Azerbaijan\\
% ...
$^{85}$ CPPM, Aix-Marseille Universit{\'e} and CNRS/IN2P3, Marseille, France\\
% ...
$^{90}$ Department of Physics and Astronomy, Michigan State University, East Lansing MI, United States of America\\
% ...
$^{107}$ Nikhef National Institute for Subatomic Physics and University of Amsterdam, Amsterdam, Netherlands\\
% ...
$^{113}$ Homer L. Dodge Department of Physics and Astronomy, University of Oklahoma, Norman OK, United States of America\\
% ...
$^{116}$ Center for High Energy Physics, University of Oregon, Eugene OR, United States of America\\
% ...
$^{151}$ Institute of Physics, Academia Sinica, Taipei, Taiwan\\
$^{152}$ Department of Physics, Technion: Israel Institute of Technology, Haifa, Israel\\
$^{153}$ Raymond and Beverly Sackler School of Physics and Astronomy, Tel Aviv University, Tel Aviv, Israel\\
% ...
$^{158}$ Department of Physics, University of Toronto, Toronto ON, Canada\\
% ...
\end{flushleft}

\end{document}